\documentclass[12pt, preprint]{aastex}

\newcommand{\WMAP}{\textsl{WMAP}}

\providecommand{\hii}{{\ion{H}{2}}}

\shorttitle{\WMAP\ 5 Year Galactic Foreground Emission}
\shortauthors{Gold et al.}

\begin{document}
\title{Five-Year Wilkinson Microwave Anisotropy Probe (\WMAP\altaffilmark{1}) Observations: \\
Galactic Foreground Emission}
\author{
B. Gold\altaffilmark{2},
C. L. Bennett\altaffilmark{2},
R. S. Hill\altaffilmark{3}, 
G. Hinshaw\altaffilmark{4}, 
N. Odegard\altaffilmark{3}, 
L. Page\altaffilmark{5}, 
D. N. Spergel\altaffilmark{6,7}, 
J. L. Weiland\altaffilmark{3},   
J. Dunkley\altaffilmark{5,7,8}, 
M. Halpern\altaffilmark{9}, 
N. Jarosik\altaffilmark{5}, 
A. Kogut\altaffilmark{4}, 
E. Komatsu\altaffilmark{10}, 
D. Larson\altaffilmark{2}, 
S. S. Meyer\altaffilmark{11},
M. R. Nolta\altaffilmark{12}, 
E. Wollack\altaffilmark{4}, 
and E. L. Wright\altaffilmark{13}
}
\altaffiltext{1}{{\WMAP} is the result of a partnership between 
                 Princeton  University and NASA's Goddard Space Flight Center. 
                 Scientific guidance is provided by the 
                 {\WMAP} Science Team.}
\altaffiltext{2}{Dept. of Physics \& Astronomy, %
            The Johns Hopkins University, 3400 N. Charles St., %
	    Baltimore, MD  21218-2686}
\altaffiltext{3}{Adnet Systems, Inc., 7515 Mission Dr., Suite A1C1 Lanham, Maryland 20706}
\altaffiltext{4}{Code 665, NASA/Goddard Space Flight Center, %
            Greenbelt, MD 20771}
\altaffiltext{5}{Dept. of Physics, Jadwin Hall, %
            Princeton University, Princeton, NJ 08544-0708}
\altaffiltext{6}{Dept. of Astrophysical Sciences, %
            Peyton Hall, Princeton University, Princeton, NJ 08544-1001}
\altaffiltext{7}{Princeton Center for Theoretical Physics, Princeton University, Princeton, NJ 08544}
\altaffiltext{8}{Astrophysics, University of Oxford, Keble Road, Oxford, OX1 3RH, UK}
\altaffiltext{9}{Dept. of Physics and Astronomy, University of %
            British Columbia, Vancouver, BC  Canada V6T 1Z1}
\altaffiltext{10}{Univ. of Texas, Austin, Dept. of Astronomy, %
            2511 Speedway, RLM 15.306, Austin, TX 78712}
\altaffiltext{11}{Depts. of Astrophysics and Physics, KICP and EFI, %
            University of Chicago, Chicago, IL 60637}
\altaffiltext{12}{Canadian Institute for Theoretical Astrophysics, %
            60 St. George St, University of Toronto, %
	    Toronto, ON  Canada M5S 3H8}
\altaffiltext{13}{PAB 3-909, UCLA Physics \& Astronomy, PO Box 951547, %
            Los Angeles, CA 90095--1547}

\setcounter{footnote}{13}

\email{bgold@pha.jhu.edu}

\begin{abstract}
We present a new estimate of foreground emission in the \WMAP\ data, using a Markov chain Monte Carlo (MCMC) method.  The new technique delivers maps of each foreground component for a variety of foreground models with estimates of the uncertainty of each foreground component, and it provides an overall goodness-of-fit estimate.  The resulting foreground maps are in broad agreement with those from previous techniques used both within the collaboration and by other authors.

We find that for \WMAP\ data, a simple model with power-law synchrotron, free-free, and thermal dust components fits 90\% of the sky with a reduced $\chi^2_\nu$ of 1.14.  However, the model does not work well inside the Galactic plane.  The addition of either synchrotron steepening or a modified spinning dust model improves the fit.  
This component may account for up to 14\% of the total flux at Ka-band (33 GHz).
We find no evidence for foreground contamination of the CMB temperature map in the 85\% of the sky used for cosmological analysis.
\end{abstract}

\slugcomment{Revised version, accepted by ApJS}
\maketitle


\section{Introduction}
The Wilkinson Microwave Anisotropy Probe (\WMAP) produces temperature and linear polarization radio maps at five frequencies with $1\degr$ or better resolution and tightly constrained systematic errors.  The frequency bands are centered on 22, 33, 41, 61, and 94 GHz; denoted K, Ka, Q, V, and W, respectively (see \citealt{page/etal:2003b} for details). While designed to measure the cosmic microwave background (CMB) radiation it also observes the large-scale structure of our Galaxy at angular scales and frequencies that are relatively unexplored.
Study of our own Galaxy has had a significant effect 
on our understanding of galaxies in general.

Radio emission from galaxies is generally understood as arising from three effects: ``non-thermal'' synchrotron emission from relativistic electrons spiraling in large-scale 
 magnetic fields, ``thermal'' free-free emission from non-relativistic electron-ion interactions, and emission from vibrational modes of thermal dust grains. At lower radio frequencies the synchrotron emission is usually dominant, with flux decreasing at higher frequencies approximately according to a power law\footnote{In this paper we use the notation that flux density is $S \sim \nu^\alpha$ and antenna temperature is $T \sim \nu^\beta$, with the spectral indices related by $\beta = \alpha-2$.  Unless otherwise noted, results will be expressed in antenna temperature.  For the pixel size most commonly used in this work ($0\fdg92 \times 0\fdg92$), the conversion from antenna temperature to flux is approximately $4.0 (\nu/22.5\textrm{ GHz})^2\textrm{ Jy mK}^{-1}$ \citep{page/etal:2003b}.}
 ($\beta \approx -3$).
Free-free emission has a flux that is nearly constant with frequency ($\beta \approx -2.1$),
so free-free emission becomes relatively more important than synchrotron at higher frequencies. Typically the crossover frequency is near 60 GHz at higher latitudes, but can be 20 GHz or lower in specific regions in the Galactic plane. Frequencies above $\sim$60 GHz begin to probe the tail ($\beta \approx 2)$ of vibrational dust emission, which is dominant around 90 GHz.  In addition to these three foregrounds, much recent work has focused on the possibility of significant emission from rapidly rotating dust grains; this emission is thought to peak somewhere in the 10--30 GHz range and fall off roughly exponentially at higher frequencies.

The spectral behavior for diffuse foregrounds is of great interest.  The spectrum for synchrotron radiation follows the energy distribution for high-energy electrons, which is not a pure power law.  The highest energy electrons lose energy more quickly and thus are reduced in regions where they have not been replenished.  Such energy loss shows up as a gradual steepening ($d\beta / d\nu < 0$) in the power law index by about $0.5$ at frequencies above $10$--$100$ MHz.  Further, while the overall index as extrapolated from lower frequencies is $\beta \approx -2.7$ \citep{reich/reich:1988,lawson/etal:1987,reich/reich/testori:2004}, higher frequencies may preferentially sample more energetic electron populations and thus have a flatter index ($\beta \approx -2.5$) \citep{bennett/etal:2003c}.  Observations of both discrete sources \citep{green:1988, green/scheuer:1992} and external galaxies \citep{hummel/dahlem/vanderhulst:1991} show a wide variety of synchrotron behavior.
Free-free emission also does not follow a strict power law, but the physics is well understood and the  variation of the power-law index over \WMAP's bands is so small that it can be neglected.
Finally, the Rayleigh-Jeans tail for vibrational dust emission (i.e. below $\!\!\sim$100 GHz) has never before been accurately measured and the relevant material properties of the dust grains themselves are not fully understood \citep{1994Natur.372..243A, 2007A&A...468..171M}.

The main focus in this work is on foreground emission.  
Section \ref{sec:fiveyearfits} describes updates to masks and foreground-fitting procedures used in previous \WMAP\ analyses \citep{bennett/etal:2003c, hinshaw/etal:2007}.  A new method to explicitly marginalize over foregrounds for the low multipole analysis is described in a companion paper \citep{dunkley/etal:prep}.
A new fitting process is described in Section \ref{sec:mcmc}, which has the following  features:
\begin{itemize}
\item The fitting is entirely in real-space with no spherical harmonic decomposition for any component.
\item The spectral indices of the synchrotron and dust emission are not generally assumed to be constant and are allowed to vary down to the scale of the fit (approximately one square degree).
\item The fit includes the CMB and automatically generates the full likelihood (including covariance) for all foreground parameters.
\item The polarization data are included and fit simultaneously with the total intensity data.
\end{itemize}
This is similar to the technique of \cite{eriksen/etal:prep}, however we fit the CMB in pixel-space, use less smoothing on the maps, and attempt to obtain more information about individual foregrounds.

The results of the fit are described in Section \ref{sec:results}.
While the fitting technique used here delivers a CMB map with error-bars, the map itself has not proved to be any better for cosmological analysis and so far has been used only as a check.  Implications of the fit are discussed in Section \ref{sec:discussion}.  \WMAP's cosmological results do not depend on the fitting process used here. 

\section{Five-year Foreground Fits\label{sec:fiveyearfits}}

\subsection{Masks}
The diffuse foreground masks are updated for the
five-year data analysis.  The primary reason is to mask
out free-free emission in the areas of the Gum Nebula
and $\rho$ Oph, while keeping a simple method that applies
to the whole sky rather than being \emph{ad hoc} for
these regions.  

The new masks are based on three-year public 
\WMAP\ data products\footnote{The new masks were based on three-year data because they were needed before the five-year maps could be finalized. The masks are made from flux cuts at high signal-to-noise on smoothed maps, thus the difference between basing the masks on three-year versus five-year data is minimal.  This was verified explicitly once the five-year maps were finalized.}, specifically the three-year
K and Q band-average maps smoothed to one-degree resolution.
These maps are converted to foreground-only maps
by subtracting the 
three-year Internal Linear Combination (ILC) map.  A cumulative histogram is made
of the pixels in each foreground map, which serves
as a lookup table to find a flux level used
to define a cut over the desired percentage of the sky.

Cuts are made at intervals of $5\%$ in the proportion of
sky admitted by the resulting mask.
The K and Q band cuts at each percentage level are combined.
Resulting masks are inspected and compared with the
masks used in the one and three-year \WMAP\ data analyses.
We replace the old Kp2 mask with the combined K and
Q $85\%$ masks.  This is the nominal mask for temperature data analysis and is denoted KQ85.
We replace the old Kp0 mask with the
combined K and Q $75\%$ masks (KQ75).

\begin{figure}
 \epsscale{0.7}
 \plotone{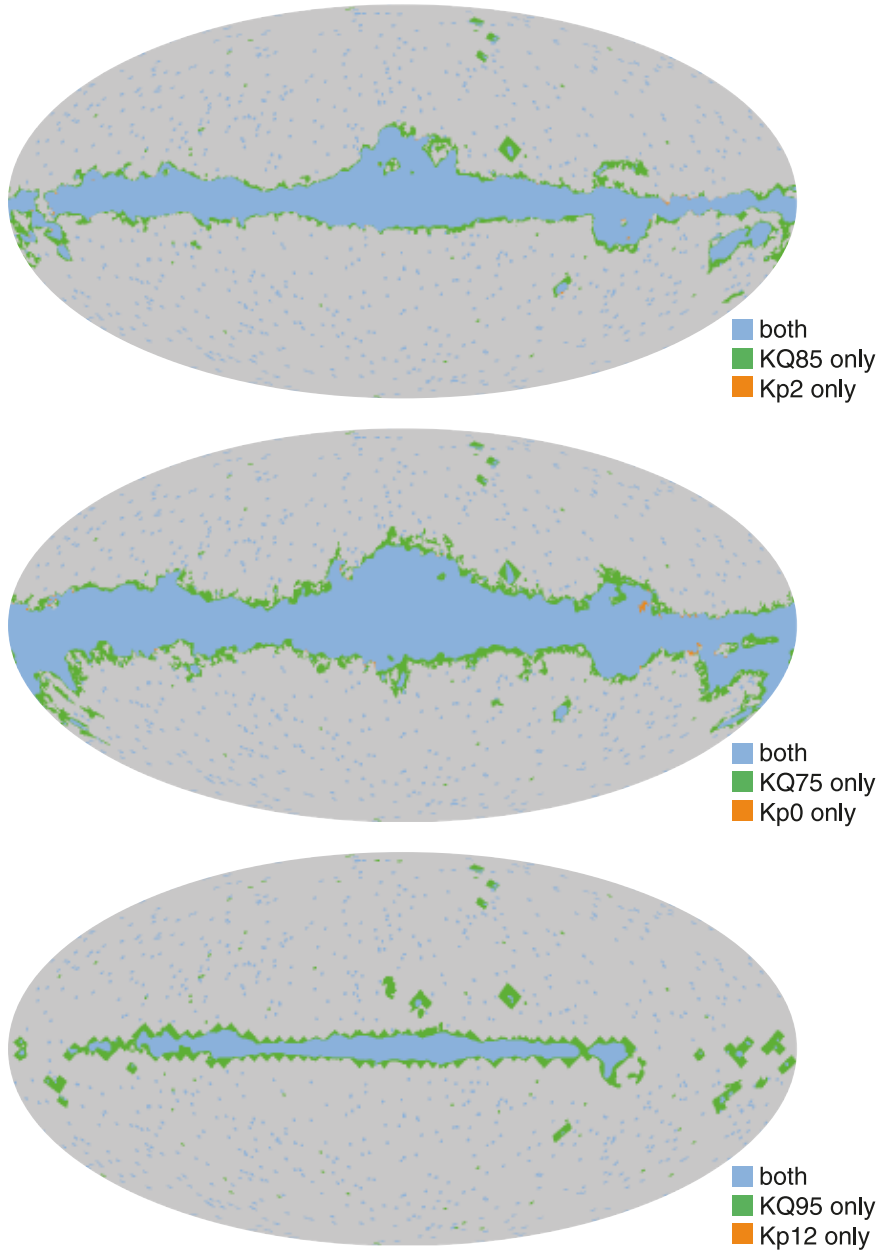}
 \caption{\label{fig:maskmap}Comparison maps of the five-year masks versus the three-year masks.
 The new masks cover slightly more of the Galactic plane and cover more regions with low synchrotron but high free-free emission.  The diamond-shaped features arise because the new processing mask has been defined to correspond to low-resolution ($N_\mathrm{side}$=16) pixels, so that the same processing mask can be used at all resolutions.  \emph{Top:} comparison of KQ85 with the three-year Kp2 mask.  \emph{Middle:} comparison of KQ75 with the three-year Kp0 mask. \emph{Bottom:} comparison of KQ95 with the three-year Kp12 mask. }
\end{figure}

Each of the chosen masks is further processed by omitting
any masked ``islands'' containing fewer than
500 pixels at HEALPix \citep{gorski/etal:2004} $N_\mathrm{side}$ of 512.  
Each mask is then combined with a point source mask, which has been updated from that described in \cite{bennett/etal:2003c} and \cite{hinshaw/etal:2007} to include 32 newly detected sources from a preliminary version of the \WMAP\ five-year point source catalog. Six sources in the final five-year catalog are not included; these are relatively weak, with fluxes of 1 Jy or lower in all \WMAP\ bands.
The last step combines each
mask with the five-year processing cut used to omit
the Galactic plane from the mapmaking.
A comparison of old and new masks is shown in Figure \ref{fig:maskmap}.

The three-year
polarization mask was based on a cut in K-band polarized
intensity combined with a model of the dust component
\citep{page/etal:2007}.
The five-year polarization analysis mask is the same
as the three-year version, with the exception that it is
combined with the five-year processing cut.  

The MCMC fit described below uses a version of the combined K and Q $95\%$ mask (denoted KQ95, and which is similar to the old Kp12 mask) to distinguish ``inside'' from ``outside'' the Galactic plane.  The mask was enlarged to account for smoothing, leaving approximately $91\%$ of the sky.  

\subsection{Internal Linear Combination Method}
The Internal Linear Combination (ILC) method is used to produce a CMB map that is independent of both external data and assumptions about foreground emission.  By construction, it leaves unchanged the component that has the spectrum of the CMB and acts as a foreground fit by filtering out the combined spectral shape that causes the most variance in the data.
As a minimum variance method the ILC is guaranteed to produce a map with good statistical properties, but the level of remaining contamination can be difficult to assess.

\begin{deluxetable}{crrrrr}
      \tablewidth{0pt}
      \tablecaption{ILC coefficients per region\tablenotemark{a}\label{tab:ilc}}
      \startdata
      \hline\hline
      Region & 
       	\multicolumn{1}{c}{K-band} &
	\multicolumn{1}{c}{Ka-band}   &
	\multicolumn{1}{c}{Q-band}      &
	\multicolumn{1}{c}{V-band}  &
	\multicolumn{1}{c}{W-band}\\
      \hline
    0&      0.1336 &     -0.6457 &     -0.3768 &     2.2940  &   -0.4051\\
    1&     -0.0610 &     -0.1327 &     -0.1873  &    1.7691  &   -0.3880\\
    2&      0.0037 &    -0.2432 &    -0.3792   &   1.7956   &  -0.1768\\
    3&     -0.1104 &     0.2395  &   -0.6424  &    1.5032    &  0.0101\\
    4&     -0.0843  &    0.1271  &   -0.4584   &   0.9739 &      0.4417\\
    5&      0.1918  &   -0.7238 &    -0.4902   &   2.4844  &   -0.4622\\
    6&     -0.1052 &      0.2614 &    -0.6223  &    1.0253   &   0.4407\\
    7 &     0.0913 &    -0.3849 &    -0.6033   &   2.3288   &  -0.4319\\
    8 &     0.2208  &   -0.5436 &    -1.0938   &   3.2084  &   -0.7918\\
    9 &    -0.0922  &   -0.0695  &   -0.1810   &   1.2619  &    0.0808\\
   10&      0.1724  &   -0.9608  &    0.0350   &   2.6456  &   -0.8923\\
   11 &     0.2374   &  -0.8975  &   -0.4897   &   2.7246  &   -0.5747\\
      \enddata
      \tablenotetext{a}{The ILC temperature (in thermodynamic units) at pixel $p$ of region $n$ is $T_n(p) = \sum_{i=1}^5 \zeta_{n,i} T^i(p)$, where $\zeta$ are the coefficients above and the sum is over \WMAP's frequency bands.}
\end{deluxetable}

\begin{figure}
 \epsscale{1.0}
 \plotone{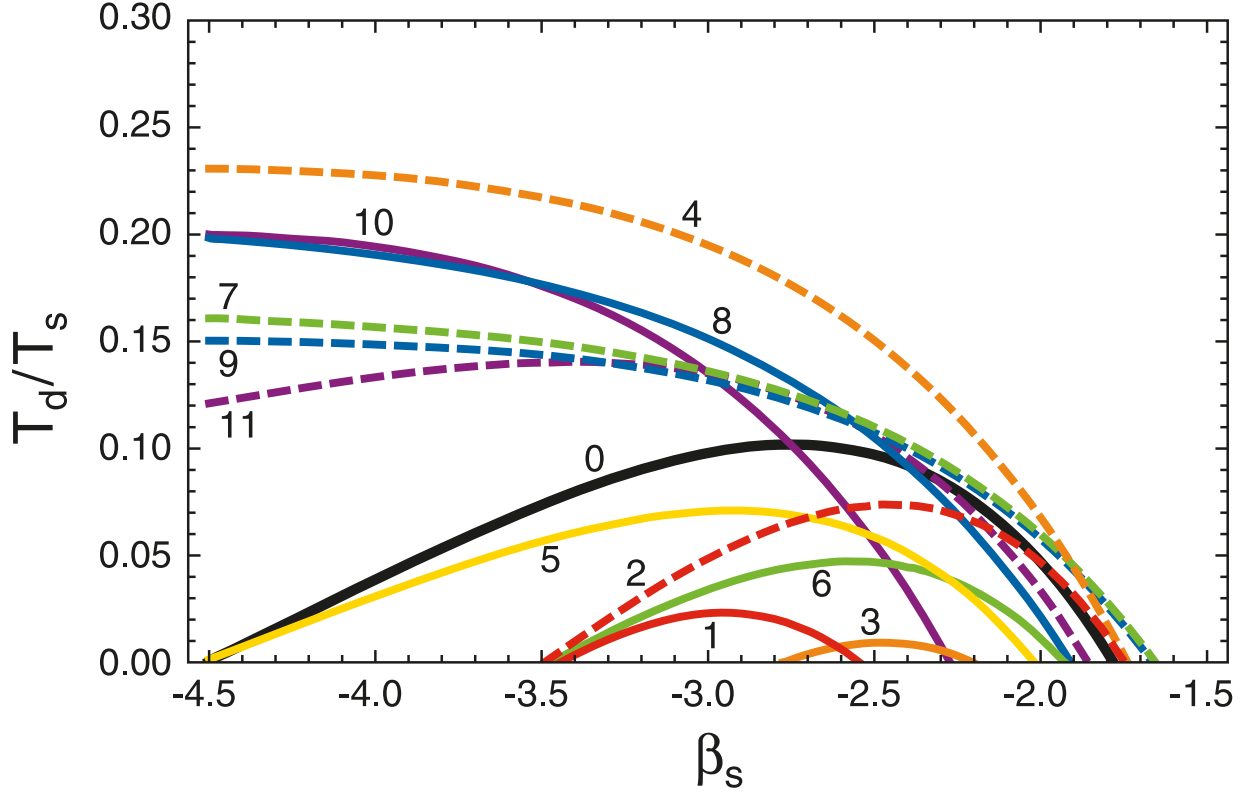}
 \caption{\label{fig:ilcfilt}A slice in parameter space of the surface nulled by the ILC coefficients, assuming a three-component foreground model with power-law spectral behavior, $T(\nu) = T_s \nu^{\beta_s} + T_f \nu^{2.14}+ T_d\nu^{\beta_d}$.  Each line is for a single ILC region, denoted by number.  The parameter space is $T_f/T_s$, $T_d/T_s$, $\beta_s$, $\beta_d$.  For this plot the x-axis is $\beta_s$ and the y-axis is $T_d/T_s$.  The parameters $T_f/T_s$ and $\beta_d$ are fixed at 0.7 and 1.8, respectively.  Each color is a different ILC region.  Despite the variety amongst ILC coefficients, they often null similar regions of parameter space.}
\end{figure}

The algorithm used to compute the \WMAP\ five-year Internal Linear
Combination map is the same as that described in the three-year analysis
\citep{hinshaw/etal:2007}. 
We retain the same number of regional subdivisions
of the sky and their spatial boundaries remain unchanged from the
previous definitions. The frequency weights for each region are somewhat
different, however, reflecting the five-year updates to the calibration and beams.
The new ILC regional coefficients are presented in Table \ref{tab:ilc}, and the map itself is
available on the LAMBDA web site\footnote{\texttt{http://lambda.gsfc.nasa.gov/}}. The coefficients describe a filter that nulls certain spectral shapes.  A slice in parameter space of the spectra nulled by the ILC is shown in Figure \ref{fig:ilcfilt}.
Differences between new CMB maps and those from the three-year release are further discussed in Section 5.

\subsection{Maximum Entropy Method}
The maximum entropy method (MEM) is a spatial and spectral fit using templates that are intended to distinguish different low-frequency emission sources.
By design, the MEM reverts to templates made from external data where \WMAP's signal is low. One of the main goals for the MEM was to use high-signal regions to investigate the spectral properties of the foregrounds. The error properties for MEM maps are complicated and the model is essentially under-constrained so there is no meaningful goodness-of-fit statistic. The MEM maps were not used for analysis of the CMB itself.  

The five-year MEM analysis is largely unchanged from the three-year analysis \citep{hinshaw/etal:2007}.  As before, the
analysis is done on sky maps smoothed to a common resolution of $1^{\circ}$ full width at half maximum in all bands.  To 
improve the signal-to-noise ratio, we now use maps degraded to HEALPix $N_\mathrm{side}=128$ pixelization instead of $N_\mathrm{side}=256$  (the pixel size 
for the former is $0.46^{\circ}$).  In the first year and three-year analyses, the logarithmic term that forces the solution to 
converge to the priors for low S/N pixels was missing a factor of $e$ \citep{cornwell/braun/briggs:1999}; this has been fixed.  The model is fit for each pixel p by minimizing 
the functional $H = A + \lambda B$ \citep{press/etal:NRIC:2e}, where $A$ is the standard $\chi^2$ of the model 
fit, and we now use $B = \sum_c T_c(p) \, \ln [T_c(p)/(e P_c(p))]$.  Here $T_c(p)$ is the model brightness of
emission component $c$ (synchrotron, free-free, dust) in pixel $p$, and $P_c(p)$ is the prior estimate of $T_c(p)$. The 
parameter $\lambda$ controls the relative weight of $A$ (the data) and $B$ (the prior information) in the fit.  An iterative 
procedure is followed that uses residuals from the fit at each iteration to adjust the spectrum of the synchrotron component 
for each pixel.  The MEM procedure was run for 11 iterations before stopping, the same as in the three-year analysis.

\begin{figure}
 \plotone{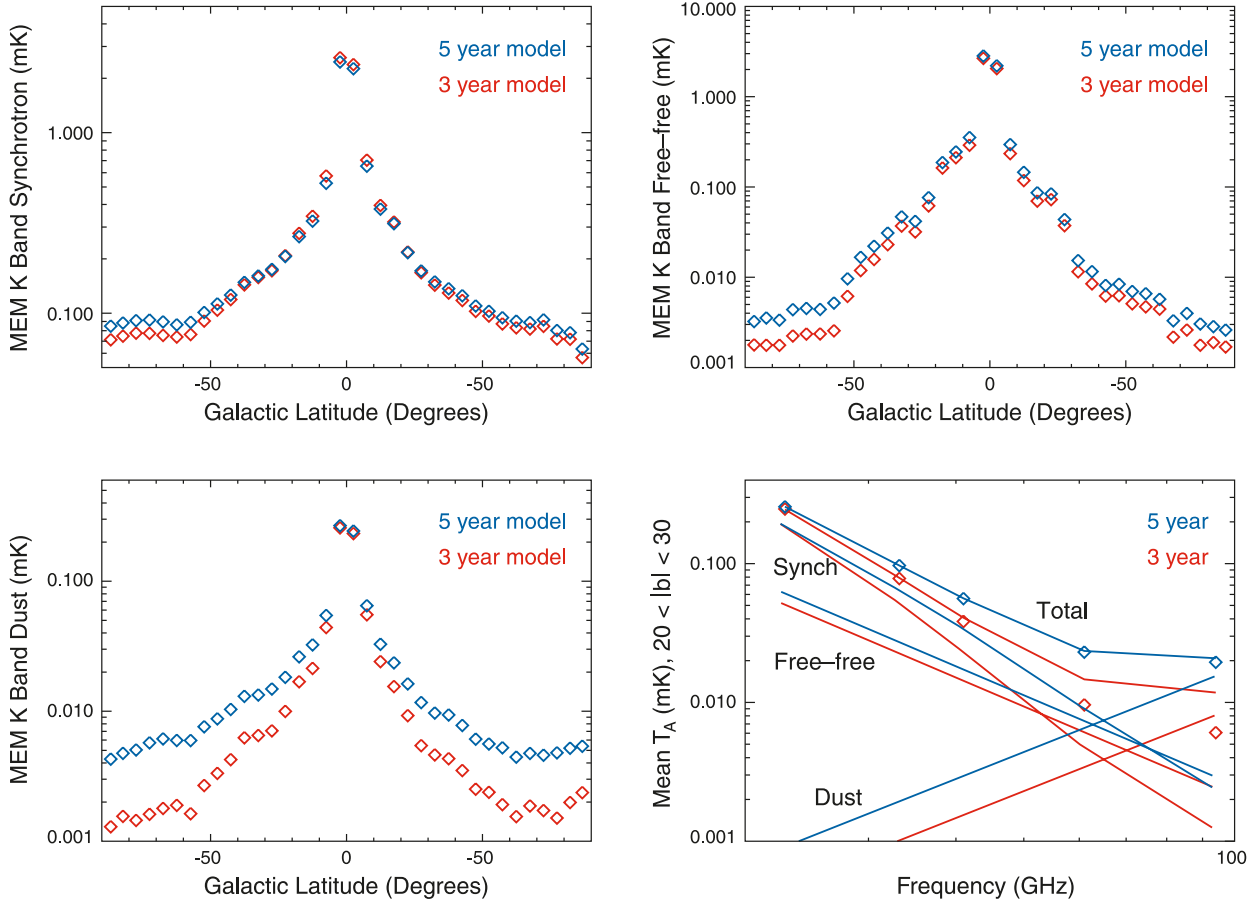}
 \caption{
 Comparison of MEM foreground modeling results from the WMAP three-year and five-year analyses.  The first three panels show latitude profiles of antenna temperature for the individual foreground model components.  The last panel compares the observed foreground emission spectrum (diamonds) with spectra of the total MEM model and the individual model components (line segments between WMAP frequencies), averaged over $20^{\circ} < \vert b \vert < 30^{\circ}$.
The differences between the three-year and five-year model results are mainly due to differences in zero levels between the three-year and five-year maps, and are consistent with the three-year year estimated error of $\sim 4 \mu$K .  The mean model brightness exceeds the mean observed brightness at the higher frequencies because the observed brightness is negative for some pixels and the model is constrained to be positive for each pixel.  This is less apparent in the five-year results because there are fewer negative pixels. \label{fig:mem_figure}}
\end{figure}

The dust and free-free spectrum coefficients are required to follow power-laws, with $\beta = +2$ for dust and $\beta=-2.14$ for free-free.
Hence any ``anomalous'' component, such as electric dipole emission from spinning dust, will be included
in the synchrotron component.  
The priors used are also unchanged, using the Haslam 408 MHz map \citep{lawson/etal:1987} for the synchrotron map, extinction-corrected H$\alpha$  \citep{finkbeiner:2003} for the free-free map, and Model 8 of \cite{finkbeiner/davis/schlegel:1999} for the dust map.
The MEM maps are available for public download on the 
LAMBDA web site.  Figure \ref{fig:mem_figure} shows a comparison of the five-year and three-year MEM foregrounds, and the spectrum of components compared to the total observed foreground spectrum for $20\degr < |b| < 30\degr$.  

Comparison of MEM results from the five-year and three-year analyses shows an increase in the model brightness of all 
foreground components at high Galactic latitudes.  The changes are mostly due to differences in the zero 
levels of the five-year and three-year maps.  The inclusion of the factor of $e$ in the MEM functional also leads to a small 
contribution.  The method of setting map zero levels has not changed since the first year analysis.  The internal linear 
combination CMB map is subtracted from the $1^{\circ}$ smoothed map in each frequency band, and the zero level is set 
such that a fit to the residual map of the form $T(\vert b \vert) = T_p \, \csc \vert b \vert + c$, over the range 
$-90^{\circ} < b < -15^{\circ}$, yields $c=0$ \citep{bennett/etal:2003c}. 
The three-year analysis procedure was done using a preliminary three-year ILC map in which the monopole was nonzero.   
Offsets of 21.1, 19.4, 19.3, 19.4, and 19.6 $\mu$K should be added to the three-year K, Ka, Q, V, and W band maps, respectively, to 
give maps that yield $\csc \vert b \vert$ fit intercepts of zero when the final three-year ILC map is subtracted.

Available foreground templates are expected to trace the distribution of foreground emission more reliably than a
$\csc \vert b \vert$ model, so template fitting has been done to check the zero levels of the five-year maps.  
Because the MEM is itself a template fit, this is essentially equivalent to fitting for the zero levels within the MEM procedure.
The five-year 
ILC map was subtracted from the five-year $1^{\circ}$ smoothed maps, and the residual map for each band was fit 
to a linear combination of synchrotron, free-free, and dust templates plus a constant offset.  Uncertainties in the
zero levels of the templates were propagated to obtain an uncertainty in the derived offset value.  For the synchrotron 
template, the 408 MHz map of \cite{haslam/etal:1982} was used with an offset of 5.9 K subtracted 
\citep{lawson/etal:1987}.  The quoted zero level uncertainty of this map is $\pm 3$ K \citep{haslam/etal:1982}.  
For the free-free template, the composite all-sky H$\alpha$ map of \cite{finkbeiner:2003} was used, with a correction 
for extinction (using the dust extinction map of \citealt{schlegel/finkbeiner/davis:1998}) assuming the dust is coextensive with the emitting gas along each line of sight \citep{bennett/etal:2003c}.
The adopted zero level uncertainty is $\pm 1$ Rayleigh, as estimated by Finkbeiner for the southern H$\alpha$ data.
For the dust template, the 94 GHz emission predicted by model 8 of \cite{finkbeiner/davis/schlegel:1999} was used.
The adopted zero level uncertainty is $\pm 0.2 \mu$K, propagated from a zero level uncertainty of $\pm 0.044$ MJy sr$^{-1}$ 
for the 100 \micron\ dust map of \cite{schlegel/finkbeiner/davis:1998}.  

Fits were done to $N_\mathrm{side}=512$ pixels that are outside 
of the combined KQ85 plus point source mask and have optical depth at H$\alpha$ less than 0.5, based on the 
\cite{schlegel/finkbeiner/davis:1998} extinction map.  This pixel selection covers 74\% of the sky.
The offsets from the fits are $-25 \pm 19$, $-5.4 \pm 6.8$, $-2.2 \pm 3.9$, $-2.2 \pm 1.5$, and $-1.5 \pm 0.7 \mu$K in 
K, Ka, Q, V, and W bands, respectively.  Thus there is no evidence for significant error in the five-year map zero 
levels as determined from the $\csc \vert b \vert$ fitting.  
For comparison, northern hemisphere $\csc|b|$ fits can be used 
to estimate uncertainties in the zero levels; 
the northern hemisphere gives offsets
of $-9.2$, $3.2$, $3.5$, $-2.5$, $-5.9$ $\mu$K for K, Ka, Q, V, and W bands, respectively,
relative to the zero levels calculated from the southern hemisphere.

\subsection{Template Cleaning}
The foreground template subtraction technique used in the five-year analysis is
unchanged from that used in the three-year release.  The method is
described in \cite{hinshaw/etal:2007} for temperature cleaning and \cite{page/etal:2007} for polarization cleaning; details are not repeated here.

In summary, for temperature cleaning a model of the foreground emission is computed from a simultaneous fit
to the five-year Q, V and W-band maps, and that model is then used to
produce foreground-reduced maps suitable for cosmological studies.  \WMAP\ has two differencing assemblies (DAs) for Q and V-bands (labelled Q1, Q2, V1, and V2) and four for W-band (labelled W1 through W4), for a total of eight maps with independent noise properties.

The model takes the form
\begin{equation}
M(\nu,p) = b_1(\nu)(T_K(p) - T_{Ka}(p)) + b_2(\nu) I_\mathrm{H\alpha}(p) + b_3(\nu) M_\mathrm{dust}(p)
\end{equation}
where $p$ indicates the pixel, the frequency dependence is entirely contained in the coefficients $b_i$, 
and the spatial templates are the \WMAP\ K-Ka temperature difference map ($T_K - T_{Ka}$), the \cite{finkbeiner:2003}
composite H$\alpha$ map with an extinction correction applied ($I_\mathrm{H\alpha}$), and the
\cite{finkbeiner/davis/schlegel:1999} dust model evaluated at 94 GHz ($M_\mathrm{dust}$).
All of these spatial templates are available on LAMBDA.

The H$\alpha$ map and dust template are based on external data and have not changed since the three-year analysis.  The first template, however, has changed slightly (at the $\sim 10$ $\mu$K level) due primarily to changes in the gain calibration since the three-year release, see Figure 5 of \cite{hinshaw/etal:prep} for details.  Because this template has contributions from both synchrotron and free-free emission, foreground parameters are a mixture of $b_1(\nu)$ and $b_2(\nu)$.  For free-free emission, the ratio of K-band radio temperature to H$\alpha$ intensity is
\begin{equation}
h_\mathrm{ff} = \frac{b_2(\nu)}{S_\mathrm{ff}(\nu) - 0.552\, b_1(\nu)}
\end{equation}
where $S_\mathrm{ff}(\nu)$ is the free-free emission spectrum converted to thermodynamic temperature units and is assumed to be a power-law with $\beta = -2.14$.
The synchrotron spectral index (relative to K-band) is found via
\begin{equation}
\beta_s = \frac{ \log\left[ 0.67\, b_1(\nu) a(\nu) \right] }{ \log(\nu/\nu_K) }
\end{equation}
where $a(\nu)$ is the conversion factor from antenna temperature to thermodynamic units.

\begin{figure}
 \epsscale{0.6}
 \plotone{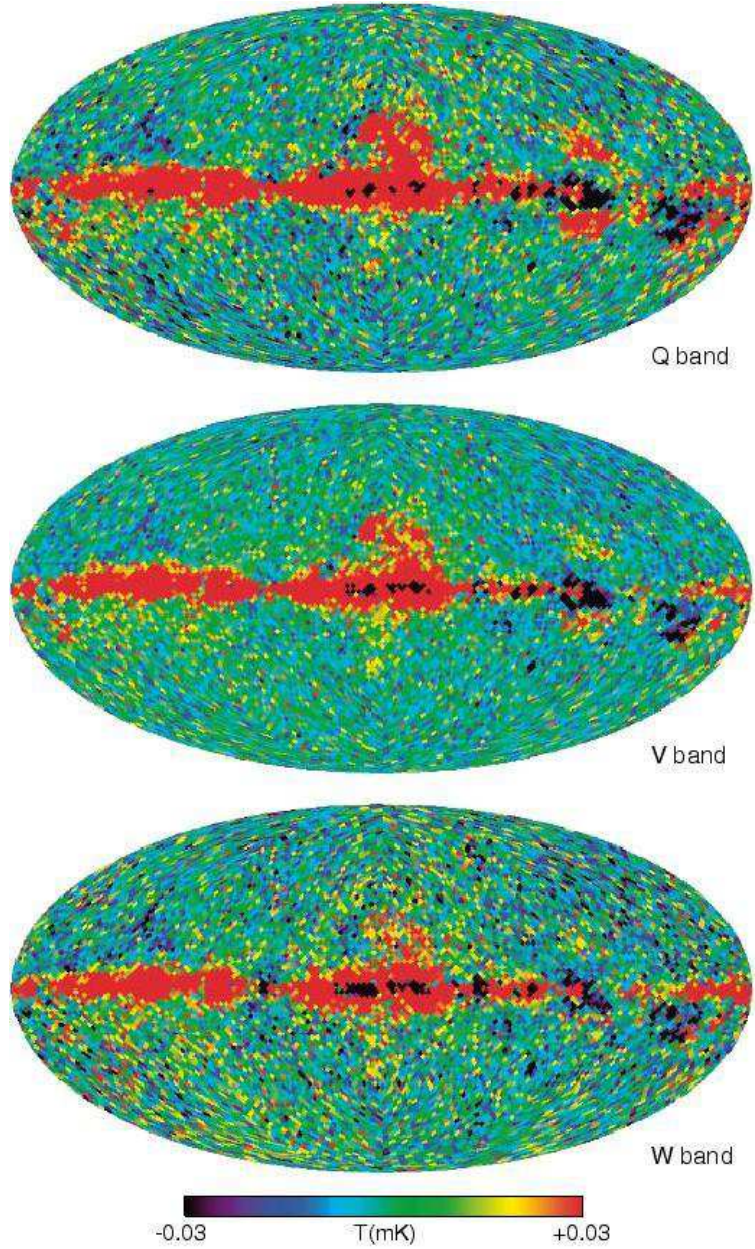}
 \caption{\label{fig:templatemaps}Five-year temperature maps with foregrounds reduced via template cleaning.  All maps have had the five-year ILC estimate for the CMB subtracted, and have been degraded to $N_\mathrm{side} = 32$.  Frequency bands shown are Q, V, and W.  Compare to Figure 10 of Hinshaw et~al. (2007).  Outside the Galactic mask, the template cleaning reduces foregrounds to $\sim 15 \mu$K or less.}
\end{figure}

The coefficients of the model fit to the five-year data are presented in
Table \ref{tab:template}.  Small changes in the five-year coefficients compared to the three-year
values (Table 5 of \citealt{hinshaw/etal:2007}) reflect the five-year updates to
absolute calibration and beam profiles.  The new template maps are shown in Figure \ref{fig:templatemaps}.

\begin{deluxetable}{cccccc}
      \tablewidth{0pt}
      \tablecaption{Template cleaning temperature coefficients \label{tab:template}}
      \startdata
      \hline\hline
      DA\tablenotemark{a} & $b_1$ & $b_2$ ($\mu$K R$^{-1}$) & $b_3$ & $\beta_s \tablenotemark{b} $ & $h_\mathrm{ff}$\tablenotemark{c}($\mu$K R$^{-1}$)\\
      \hline
      Q1 &  0.245  & 0.981 &  0.201 & -3.18  & 5.99\\
      Q2  & 0.243  & 1.009 &  0.199 & -3.22  & 6.01\\
      V1 &  0.058 &  0.666 &  0.461 & -3.44  & 6.38\\
   V2  & 0.056 &  0.647  & 0.477 & -3.43  & 6.38\\
   W1 &  0.000 &  0.398  & 1.262 & \nodata  & 6.62\\
   W2 &  0.000 &  0.393  & 1.277 & \nodata  & 6.62\\
   W3 &  0.001 &  0.398  & 1.242 & \nodata  & 6.61\\
   W4 &  0.000 &  0.395  & 1.271 & \nodata &  6.62\\
      \enddata
      \tablenotetext{a}{\WMAP\ has two differencing assemblies (DAs) for Q and V-bands and four for W-band; the high signal-to-noise in total intensity allows each DA to be fitted independently.}
      \tablenotetext{b}{Power law slope relative to K-band, as derived from $b_1$; W-band values are less than -4.}
      \tablenotetext{c}{Free-free to H$\alpha$ ratio at K-band, as derived from $b_1$ and $b_2$.  The expected value for an electron temperature of 8000 K is 11.4 $\mu$K R$^{-1}$ \citep{bennett/etal:2003c}.}
\end{deluxetable}

For polarization cleaning the maps are degraded to low resolution ($N_\mathrm{side}$ = 16).  The model has the form
\begin{equation}
  [Q(\nu,p), U(\nu,p)]_\mathrm{model} = a_1(\nu) [Q(p), U(p)]_K + a_2(\nu) [Q(p), U(p)]_\mathrm{dust}
\end{equation}
The templates used are the \WMAP\ K-band polarization for synchrotron ($[Q,U]_K$), and a low resolution version of the dust template used above with polarization direction derived from starlight measurements ($[Q,U]_\mathrm{dust}$).  While the dust polarization template maps are unchanged since the three-year release, further \WMAP\ observations have improved the signal-to-noise ratio for synchrotron polarization template maps. The coefficients of the model fit to the five-year data are in Table \ref{tab:poltemplate}.    For polarization, the template maps are assumed to have a one-to-one correspondence to foreground emission, so the spectral indices for synchrotron and dust are simply the power-law slopes of the coefficients $a_1(\nu)$ and $a_2(\nu)$.
As was the case for the three-year data, a fit fixing the synchrotron spectral index was found to have no influence on cosmological conclusions and was not used for analysis.

\begin{deluxetable}{ccccc}
\tablewidth{0pt}
\tablecaption{Template cleaning polarization coefficients\label{tab:poltemplate}}
\startdata
\hline\hline
Band & $a_1$\tablenotemark{a} & $\beta_s(\nu_K,\nu)$\tablenotemark{b} &
$a_2$\tablenotemark{a} & $\beta_d(\nu,\nu_W)\tablenotemark{b}$ \\
\hline
Ka &   0.3161  &   -3.17  &  0.0165 &    1.35 \\
Q  &  0.1765  &  -3.04 &  0.0147    &   1.85 \\
V  &  0.0595  & -2.96   &  0.0366  &   1.58 \\
W  &  0.0450   &  -2.35  &  0.0822 &   \nodata \\
\enddata
\tablenotetext{a}{The $a_i$ coefficients are dimensionless and produce model maps from templates.
}
\tablenotetext{b}{The spectral indices refer to antenna temperature.}
\end{deluxetable}


\section{Markov Chain Monte Carlo Fitting\label{sec:mcmc}}
\subsection{Description}
The analysis is carried out with band-averaged maps at each frequency, which are calibrated in antenna temperature, smoothed to a one-degree Gaussian beam, and pixelized using an $N_\mathrm{side}=64$ HEALPix grid.
This makes the fit computationally manageable and ensures that pixel-pixel correlations are small, simplifying the error description. The maps use the $\csc|b|$ fit process described above to determine the zero-point.

Next we parameterize the emission in each pixel with a physical model.  The model depends on the parameters in a non-linear way and the parameters can be highly correlated.  A Monte Carlo chain is
run for each pixel to determine the probability distribution for the parameters of the model using the Markov chain technique~\citep{gilks:MCMCIP}. Because of parameter correlations, the matrix describing the optimal step size is not diagonal. The 
starting points and initial step proposal matrices are generated using a ``best guess'' from the data.  In cases where the initial guess turns out to be poor, the fitting process is retried using the existing chain to improve the guess.  Any retries or poorly conditioned proposal matrices are flagged.  Each chain is checked for convergence using the criteria described in \citet{dunkley/etal:2005}, and any lack of convergence is also flagged.  

The basic form of the model for each pixel is
\begin{equation}
	T(\nu) = 
	T_{s} \left( \frac{\nu}{\nu_K} \right)^{\beta_s(\nu)}
	+ T_{f} \left( \frac{\nu}{\nu_K} \right)^{\beta_f} 
	+ a(\nu) T_{cmb} + T_{d} \left( \frac{\nu}{\nu_W}\right)^{\beta_d}
\end{equation}
for the antenna temperature and
\begin{equation}
	Q(\nu) = Q_{s} \left( \frac{\nu}{\nu_K} \right)^{\beta_s(\nu)} 
	+ Q_{d} \left( \frac{\nu}{\nu_W}\right)^{\beta_d} 
	+ a(\nu) Q_{cmb}
\end{equation}
\begin{equation}
	U(\nu) = U_{s} \left( \frac{\nu}{\nu_K} \right)^{\beta_s(\nu)} 
	+ U_{d} \left( \frac{\nu}{\nu_W}\right)^{\beta_d} 
	+ a(\nu) U_{cmb}
\end{equation}
for Stokes Q and U parameters. The subscripts $s,f,d$ stand for synchrotron, free-free, and dust emission, $\nu_K$ and $\nu_W$ are the effective frequencies for K and W bands ($22.5$ and $93.5$ GHz), and $a(\nu)$ accounts for the slight frequency dependence of a $2.725$ K blackbody using the thermodynamic to antenna temperature conversion factors found in \cite{bennett/etal:2003c}. 

For each pixel, the $\chi^2$ of the fit is then calculated in the standard way
\begin{equation}
	\chi^2 = \sum_\nu \mathbf{D}^T_\nu \mathbf{N}^{-1}_\nu \mathbf{D}_\nu 
\end{equation}
where $\mathbf{D}_\nu$ is the difference between the data vector ($T$, $Q$, $U$) and the model vector at each frequency.  The matrix $\mathbf{N}_\nu$ is the noise covariance matrix, and is derived directly from the $N_\mathrm{obs}$ maps (rebinned from an $N_\mathrm{side}$ of  512 to 64), with minor modifications discussed below.

Not all parameters in the model are free to vary; $a(\nu)$ and $\beta_f$ are fixed by known physics, 
the $Q$ and $U$ parameters for foregrounds are related using K band as a template for the polarization angle, and for most of the following $\beta_s(\nu)$ is assumed to be constant with frequency (though
allowed to vary spatially). The free-free index was fixed at $\beta_f = -2.14$; 
typical variation in this value at \WMAP\ frequencies is $\pm 0.015$ (\citealt{oster:1961}, see also \citealt{2006ApJ...653.1226Q, 2000ApJS..128..125I} for recent refinements) which is too small for \WMAP\ to detect.
Similarly, results of the fit were not found to depend strongly on whether $U_s$ and $U_d$ were treated as freely independent parameters. 

The ``base'' fit which allows for spatially varying synchrotron and dust spectral indices has 10 independent parameters per pixel: $T_s$, $T_f$, $T_d$, $T_{cmb}$, $\beta_s$, $\beta_d$, $Q_s$, $Q_d$, $Q_{cmb}$, and $U_{cmb}$. 
More restricted fits fixed $\beta_s$, $\beta_d$, or both.  Other fits allowed for a frequency-dependent 
$\beta_s$ by defining
\begin{equation}
	\beta_s(\nu) = \beta_s + c \ln (\nu/\nu_K)
\end{equation}
where the new parameter $c$ can introduce a gradual steepening (or shallowing).

Note that the models used here assume that polarized and unpolarized synchrotron emission have the same spectral behavior.  While this assumption appears to be safe at high latitudes, it may not be accurate for lines of sight that pass through the Galactic plane.  This is further explored in Section \ref{sec:discussion}.

\begin{figure}
 \epsscale{1.0}
 \plotone{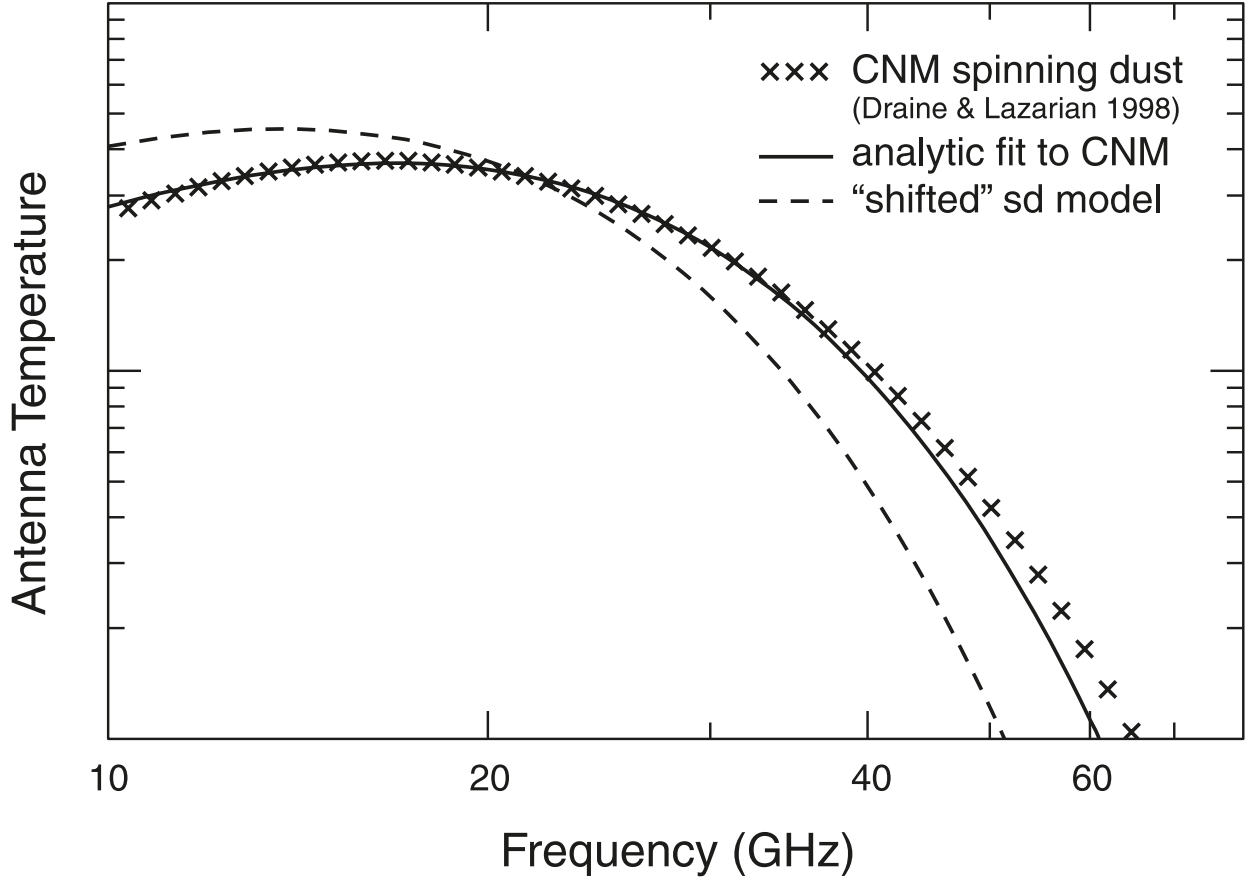}
 \caption{\label{fig:sd}The exact cold neutral medium (CNM) spinning dust spectrum as calculated by Draine \& Lazarian (1998), an analytic fit to the model, and the ad hoc ``shifted'' model which better fits radio observations in the Galactic plane.  The shifted model may represent a mix of warm neutral medium and warm ionized medium models, or another emission process entirely.  The vertical axis is in units of antenna temperature, but the overall scale is arbitrary. Agreement between the exact model and the analytic approximation is better than 5\% over the frequency range from 15 to 35 GHz.  This is smaller than the fractional error from fits that include a spinning dust component.}
\end{figure}

Finally, some fits allowed for an additional independent component, either using the exact ``cold neutral medium'' (CNM) spectrum for spinning dust \citep{draine/lazarian:1998a}, or using a generalized analytic form 
\begin{equation}
	 T_{sd}(\nu) = A_{sd} \frac{(\nu/\nu_{sd})^{\beta_d+1}}{\exp(\nu/\nu_{sd})-1}
\end{equation}
The analytic form is a modified blackbody with the amplitude, low-frequency spectral index, and turnover frequency explicitly decoupled from one another.  Plots of both the exact and ``shifted'' spectra used in the fitting process are shown in Figure \ref{fig:sd}, as well as a curve showing that the analytic form is indeed a good approximation to the numerically calculated spectrum. In practice the low-frequency spectral index is irrelevant because the desired shape for foreground fitting has $\nu_{sd}$ well below 22 GHz and is thus dominated by the exponential cutoff. While this form was motivated as an analytic approximation to spinning dust spectra it could also represent a variety of other physical sources of microwave emission.  This component was assumed to have no significant polarization.

Each chain itself is a multi-step process.  The code makes an initial guess for the best-fit parameters and runs for a burn-in period to find the region of parameter space near minimum $\chi^2$.  There is then a ``pre-chain'' to find the approximate moments of the likelihood; these moments are used to optimize the proposal distribution for the final chain.  Problems at any stage due to lack of convergence or poorly characterized parameter distributions are flagged and recorded; only rarely are more than $0.5\%$ of pixels so affected, and most problems are due to random fluctuations and disappear with longer chains.

\subsection{Tests and Sources of Error}
The Monte Carlo process (with Metropolis steps) has the advantage that it can sample the full parameter space and will converge on the likelihood even if the likelihood is non-Gaussian or unknown a priori. The disadvantage is that degeneracies in parameter space will slow the convergence, and cutting
off regions of parameter space to improve convergence can bias the results. 
The prime example of this is degeneracy between synchrotron and free-free emission amplitude.  If the synchrotron spectral index is allowed to flatten to the free-free value 
then the amplitudes of the two components become degenerate parameters, which can distort the fit.

We test this with simulated maps where the input foreground is known.
To ensure that the noise properties of the maps are well understood we used extensive simulations.  We combined the high resolution noise information with the low resolution pixel-pixel covariance in order to generate noise realizations that are as realistic as possible, which are then smoothed using the same process used for the real sky maps.
We then produce mock sky maps with CMB realizations synthesized from the {\WMAP} ``best-fit'' $\Lambda$CDM model, noise realizations from the five year noise covariance matrix, and 
foregrounds generated from a variety of models.

\begin{figure}
\epsscale{0.42}
\plotone{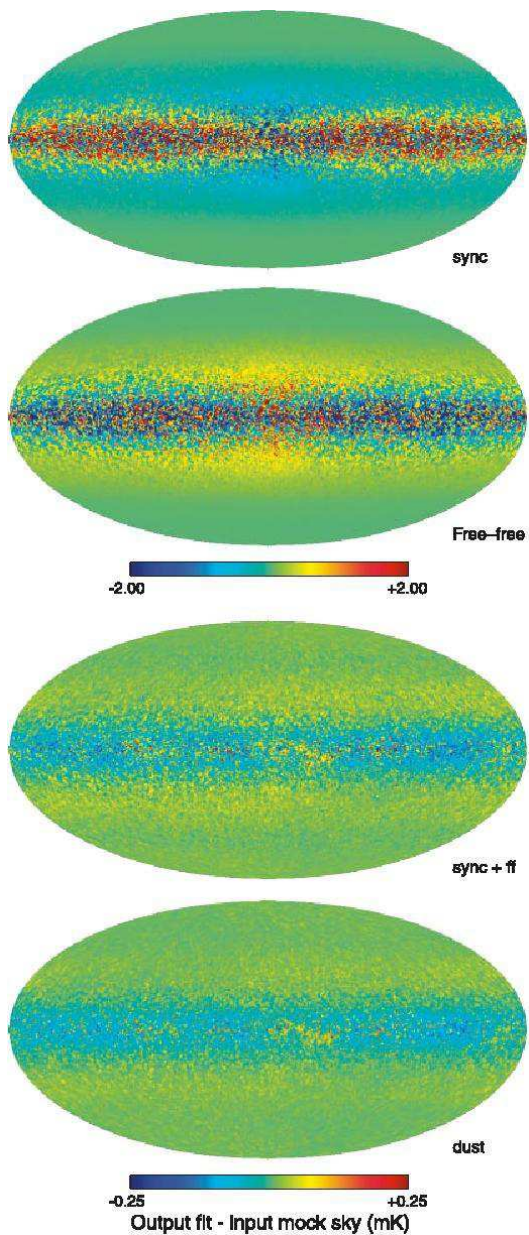}
 \caption{Difference maps between input and output foregrounds for the mock sky reconstruction. For comparison, the peak temperature for synchrotron plus free-free inputs was $17$ mK, and for the input dust the peak was $1.9$ mK. The degeneracy between synchrotron and free-free emission means their sum is much better constrained than either component individually.  The scatter is larger than the bias, which is small compared to the input signal and within the error estimate given by the fit. These degeneracy issues are also illustrated in Figure \ref{fig:mocksinglepix}. Synchrotron and free-free antenna temperatures are defined as measured at K-band, and dust as measured at W-band. \label{fig:mockfg}}
\end{figure}

\begin{figure}
\epsscale{1.0}
\plotone{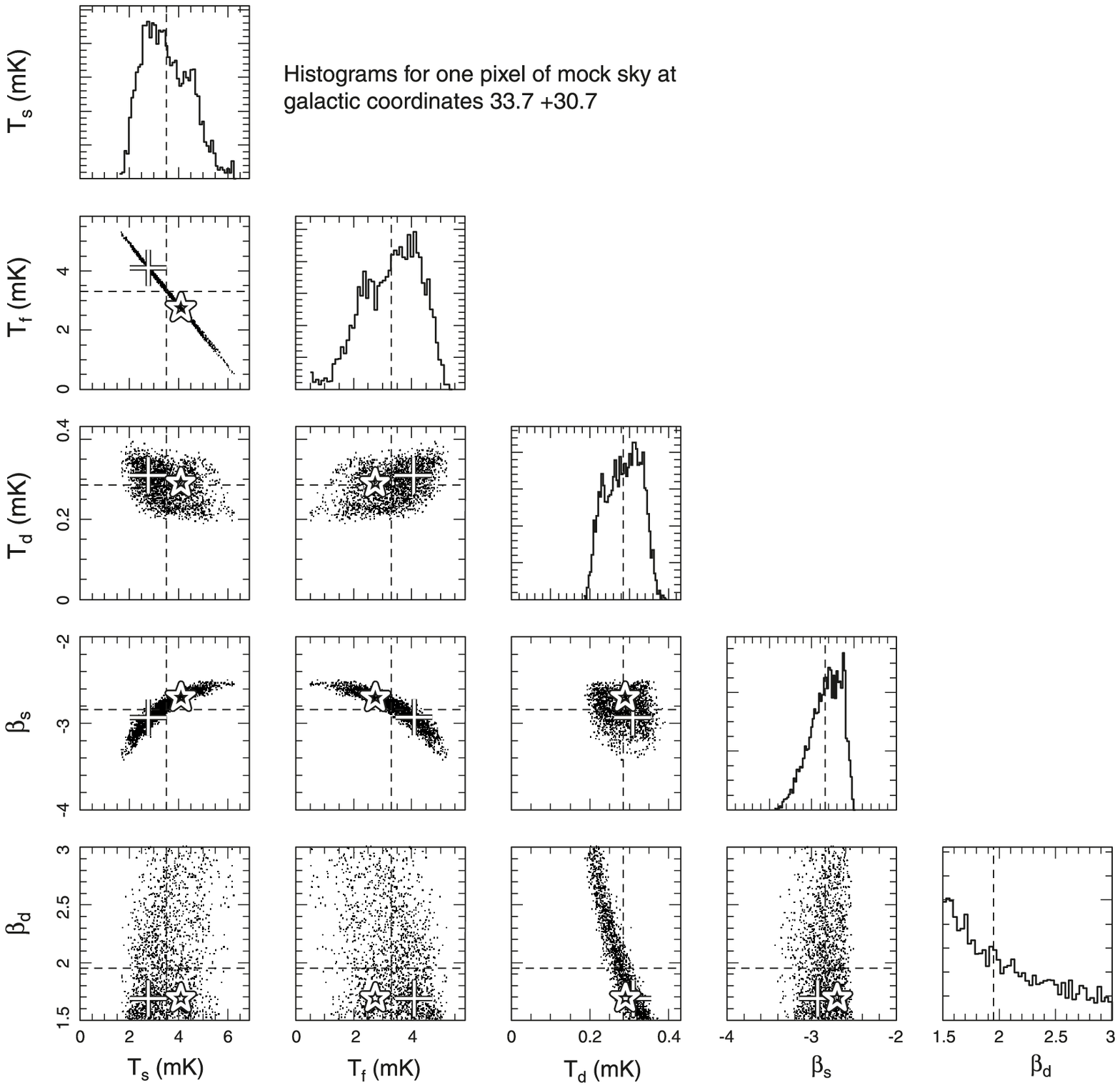}
 \caption{Histograms for foreground parameters for a single pixel of mock sky reconstruction. The mock foregrounds were designed to have the same statistical behavior as the true sky but do not match in detail.  The same pixel is examined for the true sky in Figure \ref{fig:fgsinglepix}.  Dashed lines indicate the mean of the chain, crosses indicate the best-fit point, and stars indicate the input values for the parameters.
 The strong $T_s$--$T_f$ degeneracy and the curved $T_s$--$\beta_s$ degeneracy are typical for high signal-to-noise pixels.  Other parameter correlations are minor.  Both the mean and best-fit parameters of the chain are within the expected error range from the input values.
 \label{fig:mocksinglepix}}
\end{figure}

When degeneracies exist the random fitting process tends to share the amplitude
evenly between degenerate parameters.  This can lead to biasing if the true sky does not also have equal contributions from such parameters.  This effect can be seen clearly by comparing the reconstruction of the synchrotron and free-free components (Figure \ref{fig:mockfg}).  Figure \ref{fig:mocksinglepix} shows histograms of a single pixel chain from the mock sky fit, with the input values of the parameters marked with a bold cross.

The fitting process uses the error information contained in the $N_\mathrm{obs}$ maps which includes covariance between the $Q$ and $U$ Stokes parameters within a pixel, but covariance between
pixels (due to low-frequency noise or the smoothing process) is not included.  The correlation coefficient between adjacent pixels ranges from less than $0.20$ up to $0.45$ (for K-band and W-band, respectively) due to the smoothing process.  
The fit treats pixels as independent, essentially marginalizing over all other pixels when fitting, so the main effect of the correlations is to introduce similarly small pixel-pixel correlations in the $\chi^2$ values.  This has a negligible effect on the results as long as goodness-of-fit is averaged over large enough regions.

Smoothing also reduces the overall noise level, and this has been modeled through direct noise simulations and accounted for in the fit process.  The method used was to generate many realizations of simulated noise maps based on the original $N_\mathrm{obs}$ information, smooth them using the same window functions used to smooth the real data, and fit a Gaussian shape to a histogram of the result.  From this process an overall multiplicative rescaling factor was determined for each frequency band and applied to the $N_\mathrm{obs}$ files used for the final fit.  
With this correction, the $\chi^2$ per degree of freedom should be close to unity for an ideal fit; for the mock fit described above the mean (per pixel) reduced $\chi^2$ was $1.11$ with 7.2 degrees of freedom. 

Small differences in the beam solid angle from one frequency band to another can distort the inferred spectral index, especially near bright sources.  We used the Jupiter-based beam maps from \cite{hill/etal:prep} and smoothed them to a common one-degree beam, similar to how the sky maps are smoothed, and found that beam systematics at the known level can cause deviations of up to $\pm 0.1$ in the spectral index.  Errors of this type (that are multiplied by the sky signal) are included in the fit by adding $0.3\%$ of the antenna temperature to the error budget for each pixel.  This number was derived from the observed scatter one beam-width away from bright point sources.  Systematics of this type do \emph{not} average down, and so can quickly become dominant at low resolutions.

The smoothing kernel used to match the bands to one-degree resolution uses the \emph{symmetrized} beam profile, and hence does not take into account beam asymmetries.  \WMAP's observational strategy, however, symmetrizes the beam to a large extent.  \cite{page/etal:2003b} investigated the extent to which remaining beam asymmetry could affect the beam window functions and found it to be $< 1$\%.  Any effect on the maps due to beam asymmetries should be weak near the ecliptic poles and for extended emission not aligned along the plane of the ecliptic.

There remain small uncertainties of a few $\mu$K both in the true zero-level of the maps and in the dipole subtraction process.  
The offsets primarily affect foreground estimation  by changing the apparent spectral index when averaging over large, very low signal-to-noise regions.  We avoid this by explicitly de-weighting or masking pixels with weak spectral index constraints when reporting results. 
Second, as a purely pixel-based method, the MCMC foreground fitting process we use is free to produce foreground (or even CMB) maps with non-zero monopole and dipole contributions.  
The sensitivity of the fit to offsets was checked by adding offsets of 100 $\mu$K (several times larger than the error as estimated from the $\csc|b|$ fits) to the sky maps and repeating the analysis.  
No foreground component was found to change by more than 10\% for pixels where the signal-to-noise was significant.

\section{Fit Results and Comparisons\label{sec:results}}

\subsection{MCMC Fit}
\begin{deluxetable}{lcccc}
\tablewidth{0pt}
\tablecaption{Model fits to WMAP temperature and polarization data\label{tab:chi}}
\startdata
      \hline\hline
       & \# of  & \multicolumn{3}{c}{Best-fit $\chi^2_\nu$ \tablenotemark{a}} \\
      model & params & outside plane\tablenotemark{b} & inside plane\tablenotemark{b} & full sky \\
      \hline
      base & 10 & 1.14 & 2.23 & 1.24 \\
      base + Haslam & 10 & 1.14 & 2.36 & 1.26 \\
      \hline
      loose priors & 8 & 1.09 & 3.26 & 1.29 \\
      steep & 10 & 1.14 & 0.97 & 1.13 \\
      exact sd & 9 & 1.21 & 1.63 & 1.25 \\
      shifted sd & 9 & 1.24 & 1.00 & 1.22 \\
      \hline
      $\beta_s = -3.2$, $\beta_d=1.7$ & 8 & 1.16 & 4.33 & 1.45 \\
      $\beta_s = -2.6$, $\beta_d=1.7$ & 8 & 1.30 & 3.42 & 1.50 \\
      $\beta_s$ variable, $\beta_d=1.7$ & 9 & 1.16 & 2.92 & 1.32 \\
      $\beta_s$ variable, $\beta_d$ variable & 10 & 1.14 & 2.23 & 1.24 \\
\enddata
\tablenotetext{a}{Reduced $\chi^2$ averaged over pixels in the region, with effective degrees of freedom determined by the chain. The statistical errors are less than 0.01.}
\tablenotetext{b}{The mask used to define these regions is a smoothed version of the 95\% mask, the 5-year release analogue of the Kp12 mask.}
\end{deluxetable}
Each pixel fit consists of 15 data points (Stokes I, Q, and U for each of the five frequency bands) and a foreground fitting model can use from 8 to 12 parameters per pixel.
The fitting process produces a $\chi^2$ value for each pixel.  Normally the
reduced $\chi^2$ is found by dividing by the number of degrees of freedom.
However, the true number of degrees of freedom in this case is difficult to 
determine because neither the data points nor the fitting parameters are
statistically independent of one another.
Using the MCMC chain for each pixel, though, it is possible to use the
``Bayesian complexity'' (described in a cosmological context in \citealt{kunz/trotta/parkinson:2006}), defined as the difference between the average $\chi^2$ 
over the chain and the $\chi^2$ of the best fit.  This serves as a measure of the 
effective number of degrees of freedom, and can then be used to determine
the reduced $\chi^2$ per pixel.  Using the simulated skymaps described above, we have
found that the statistical behavior
of the reduced $\chi^2$ defined this way is consistent with that of a $\chi^2$
distribution.  This ``effective'' reduced $\chi^2$ is how we quantify the
goodness-of-fit in the tables and figures.

Pixels with high reduced $\chi^2$ are not being well-fit by the model.  Since such pixels are largely confined to the plane, the sky was divided into regions ``outside'' and ``inside'' the Galactic plane by using progressively smaller masks until the average ``outside'' $\chi^2$ was no longer independent of the mask.  Regions near known point sources from \cite{wright/etal:prep} are excluded from all analysis, both inside and outside of the plane, leaving $92\%$ of the full sky.  Flagged pixels are also not included; for most fits such pixels arise from poorly conditioned covariance matrices, are uniformly distributed, and make up less than $0.5\%$ of the sky.

\begin{figure}
\plotone{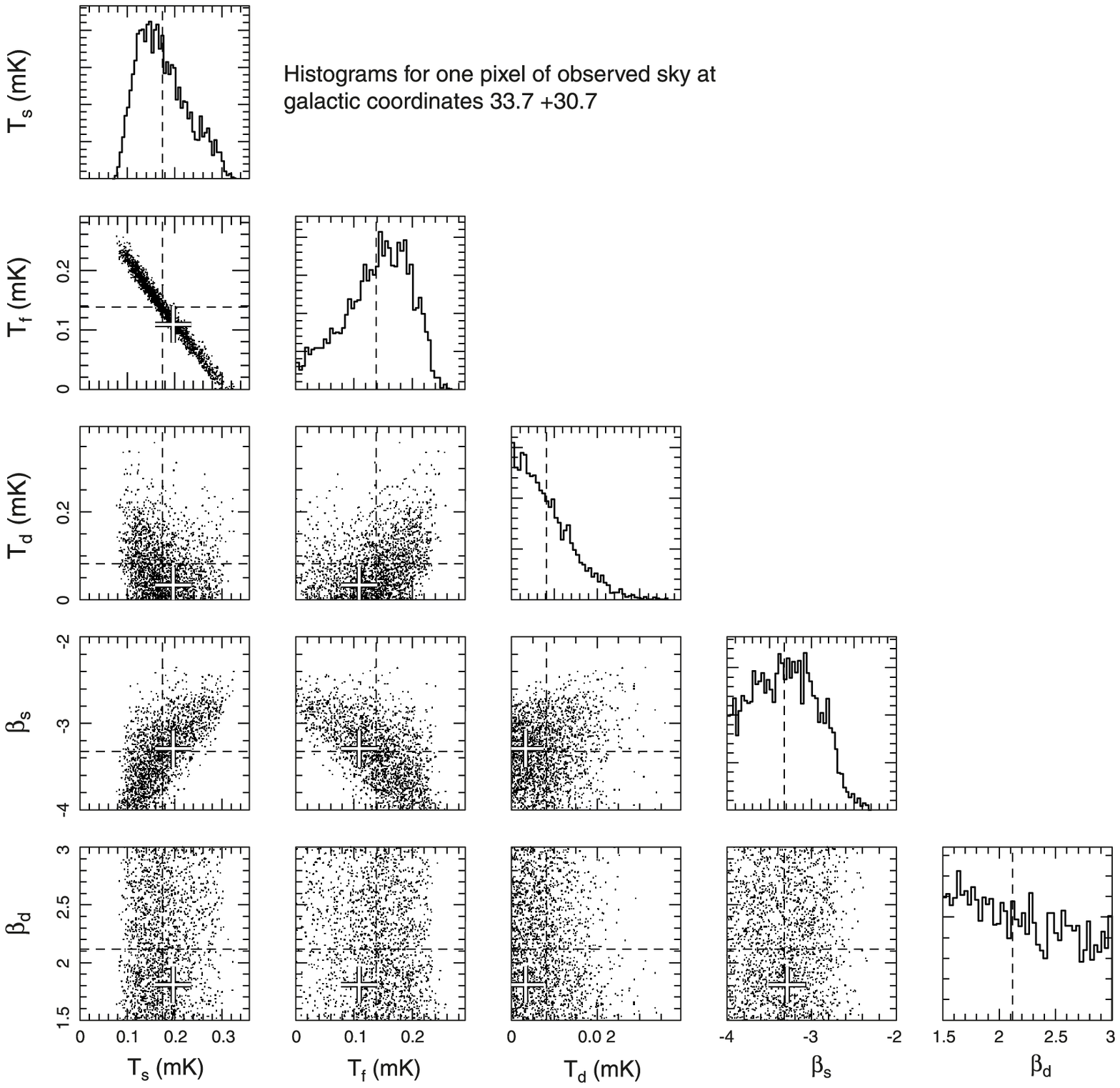}
 \caption{Histograms for foreground parameters for a single pixel of the observed sky from actual \WMAP\ data, not simulation. Dashed lines indicate the mean of the chain, and crosses indicate the best-fit point.
 The observed sky shows the same basic behavior as the simulated sky used for testing.
 \label{fig:fgsinglepix}}
\end{figure}

\begin{figure} 
\epsscale{0.7}
\plotone{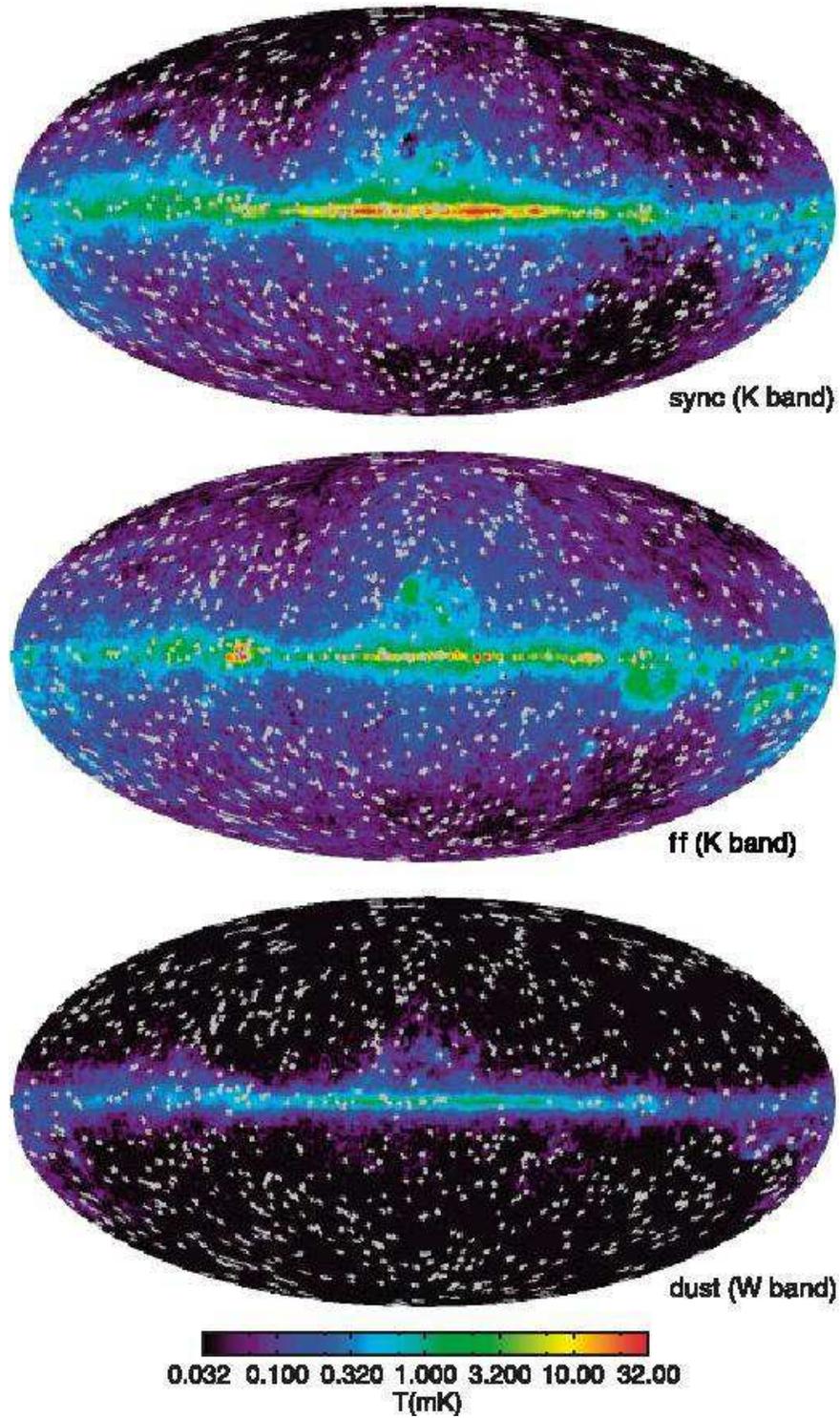}
 \caption{Temperature maps for foreground components as determined by the MCMC fitting process for the ``base'' model. Maps from other models are qualitatively similar. Synchrotron and free-free temperatures are as measured at K-band, dust is measured at W-band.  Gray pixels are those masked due to point sources or flagged as problematic.  \emph{Top:} synchrotron; \emph{middle:} free-free; \emph{bottom:} dust\label{fig:fgtemp}} 
\end{figure}

\begin{figure}
\epsscale{0.8}
\plotone{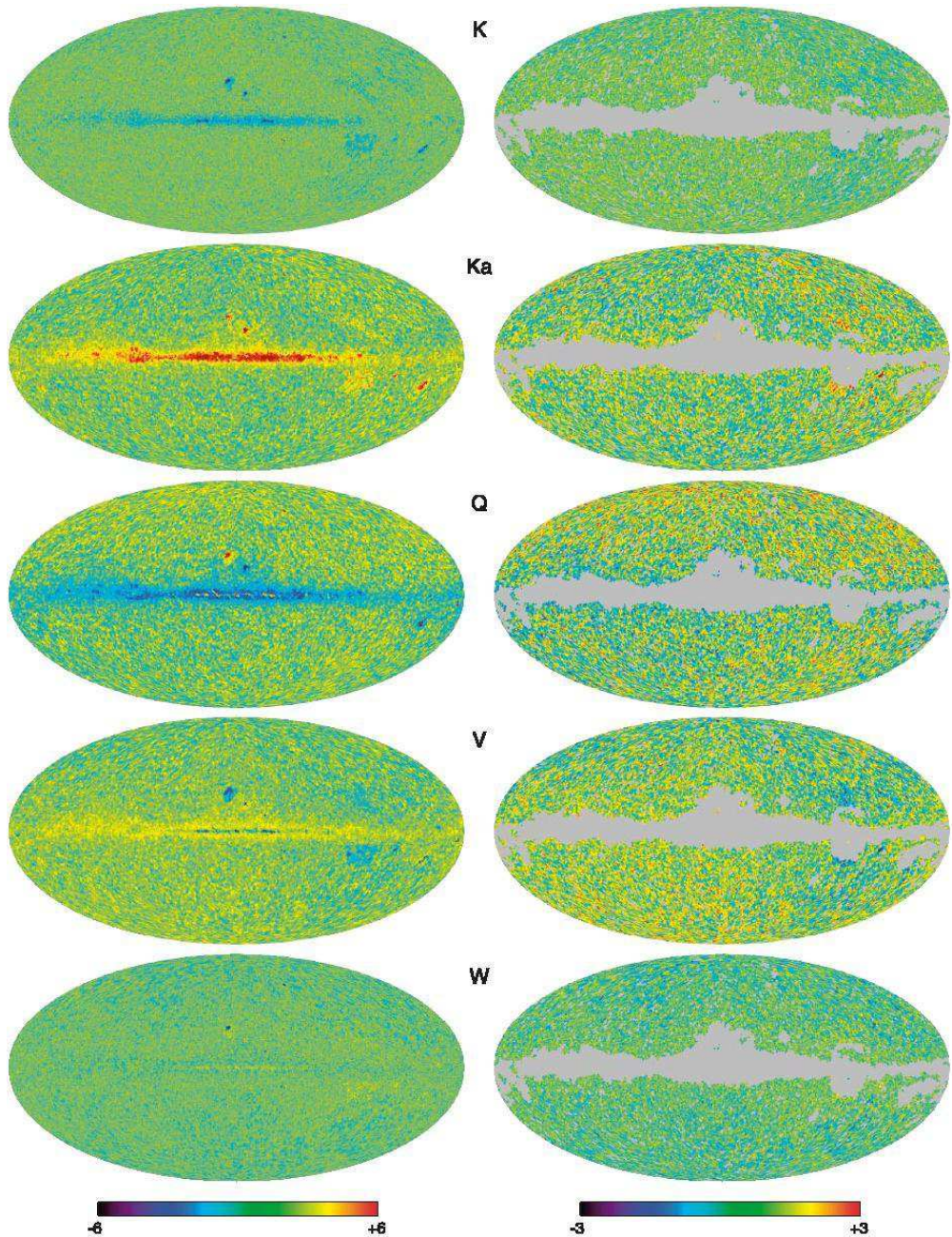}
 \caption{Residuals of the ``base'' model foreground fit subtracted from the data. The left column shows the residuals in units of noise sigma, from -6 to 6.  The right column has the 85\% mask applied and a scale of -3 to 3.  Frequency bands are (from top to bottom) K, Ka, Q, V, W.  The main feature is that the model underestimates Galactic flux in Ka band and overestimates it in Q band by a factor of 3 to 5 times the pixel noise.  Outside the analysis mask the residuals to the model are consistent with noise at the expected level.
 \label{fig:resid}}
\end{figure}

\begin{figure}
 \epsscale{1.0}
 \plotone{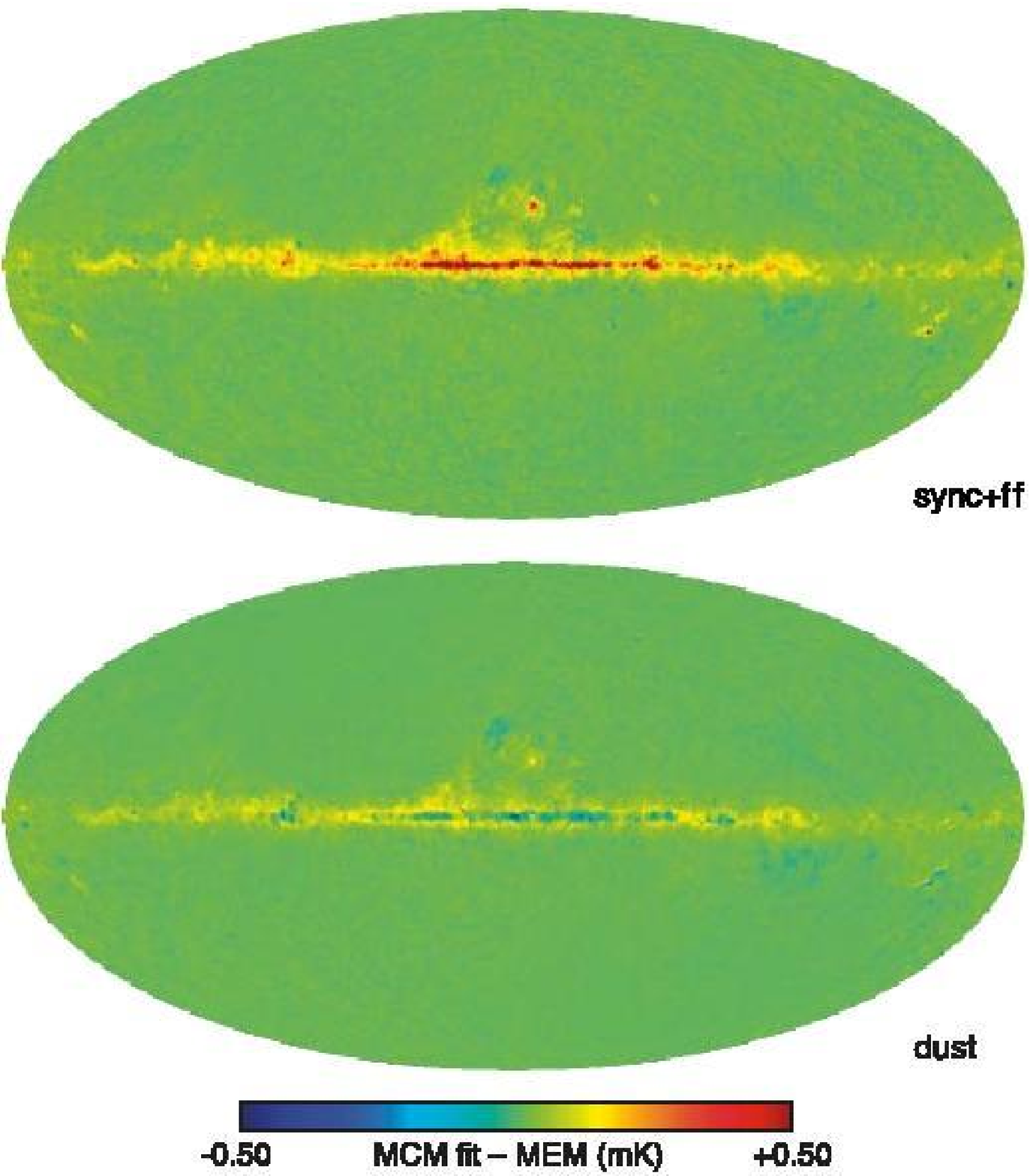}
 \caption{Difference between the MCMC fit and MEM maps.  
 The top panel is the difference (at K band) in synchrotron plus free-free emission.  Differences between these two components separately are larger due to the degeneracy direction. The bottom panel is the difference (at W band) in the dust maps. The differences are roughly one percent of total emission at K band and a few percent at W band.
 \label{fig:memdiff}}
\end{figure}

The $\chi^2_\nu$ results of several fits are shown in Table \ref{tab:chi}.  Histograms for a single pixel chain are shown in Figure \ref{fig:fgsinglepix}.  The mock sky simulations appear to capture the basic behavior of the parameter correlations.  Detailed comparisons of specific results are in the subsections below.  Figure \ref{fig:fgtemp} shows the basic results and Figure \ref{fig:resid} shows the temperature residuals of the best-fit base model subtracted from the data, in units of one-sigma of noise.  Figure \ref{fig:memdiff} shows the difference between the MCMC fit and the five-year MEM fit.
The overall fit outside the complex and troublesome Galactic plane region approaches a $\chi^2_\nu$  of 1.14 and the residuals are mostly randomly distributed, which suggests that the overall fit works reasonably well and that the noise properties have been described properly.

The ``base'' model uses the 10 parameters described above.  Another fit is done including data from 408 MHz \citep{haslam/etal:1981}, assuming 10\% calibration errors.  As a check, a ``loose priors'' fit is done which allows foreground temperatures to become negative.  For this fit to converge the spectral indices must be fixed, however, so it only uses 8 parameters.  The ``steep'' model fixes the dust spectral index but allows for a synchrotron steepening parameter $c$ as described above.  The ``exact sd'' model uses the CNM spinning dust spectrum for an additional foreground, whereas the ``shifted sd'' model uses the generalized spectrum with $\nu_\mathrm{sd} = 4.9$ GHz.  In both cases, the synchrotron and dust spectral indices are fixed to be the same at each pixel (again for convergence reasons).
Finally, several fits were done with fixed spectral indices to examine the effect of using different values, shown as the last part of Table \ref{tab:chi}.  The last entry of the table is the ``base'' model, repeated for ease of comparison. 

\subsection{Overall Foreground Features}

\begin{figure} 
\epsscale{0.7}
\plotone{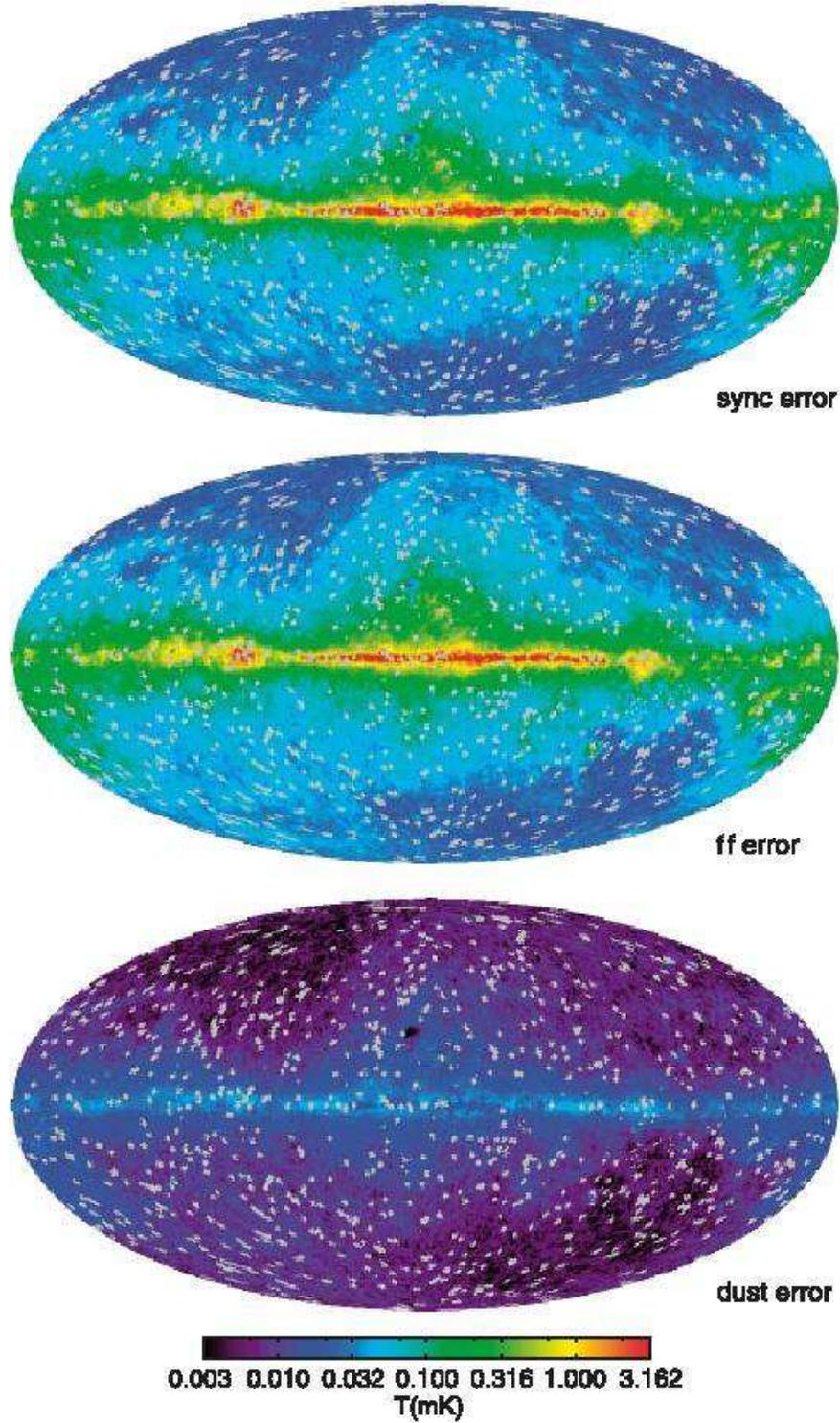}
 \caption{Error maps for foreground components as determined from the marginalized variance given by the MCMC fitting process. Error maps for synchrotron and free-free emission are similar due to the parameter degeneracy between them.  Synchrotron and free-free temperatures are as measured at K-band, dust is measured at W-band.  Gray pixels are those masked due to point sources or flagged as problematic. \emph{Top:} synchrotron; \emph{middle:} free-free; \emph{bottom:} dust \label{fig:fgerr}} 
\end{figure}

\begin{figure} 
\epsscale{0.8}
\plotone{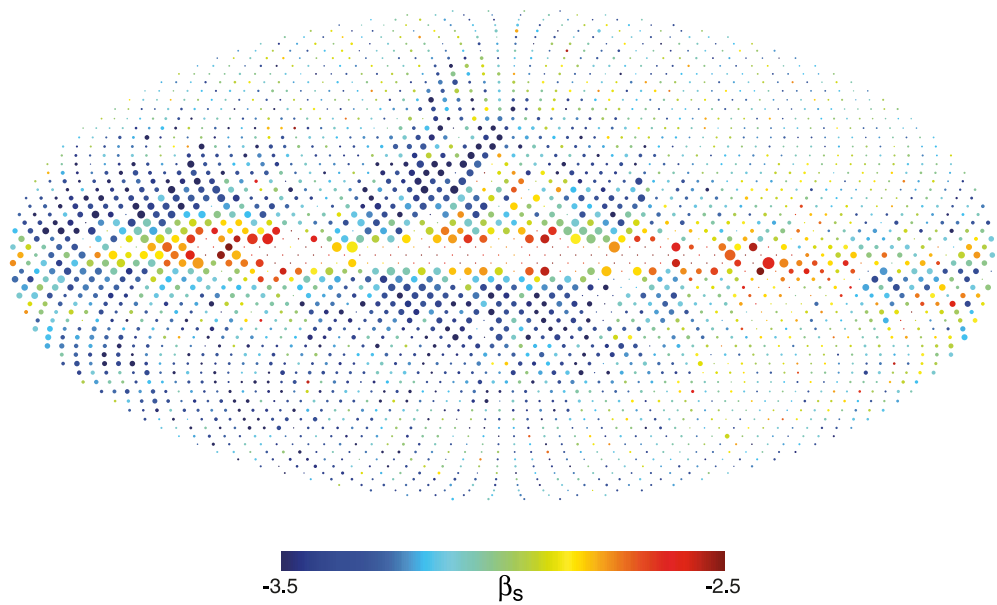}
 \caption{Map of synchrotron spectral index for the ``base'' fit, binned to $N_\mathrm{side} = 16$.  Color shows the value of the spectral index, and circle area indicates the weight ${\sigma_\beta}^{-2}$ given by the fit.  Pixels with $\chi_\nu^2 > 2$ were explicitly de-weighted.\label{fig:fgbeta}}
 \end{figure}

\begin{figure}
\epsscale{1.0}
\plotone{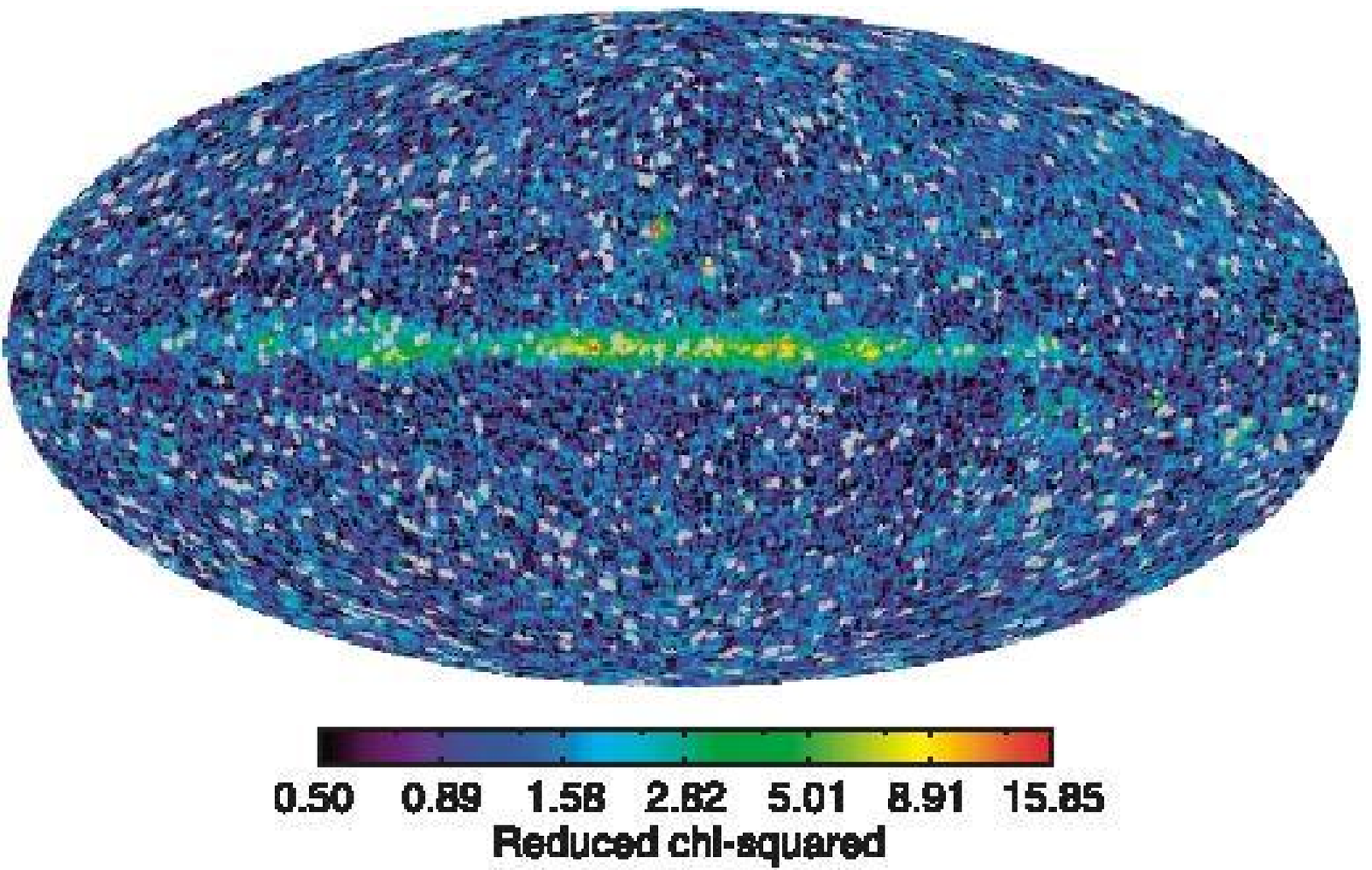}
 \caption{Map of the reduced $\chi^2$ per pixel achieved by the ``base'' fit.  Except for a slight excess of high $\chi^2$ values from the plane, the overall statistical distribution is that of an ideal $\chi^2$ distribution.
 	\label{fig:chimap}}
\end{figure}

Figures \ref{fig:fgtemp}--\ref{fig:chimap} show maps of the results from the ``base'' fit.  Figures \ref{fig:fgtemp} and \ref{fig:fgerr} show the three foregrounds themselves and their errorbars as determined from the parameter variance in the Markov chains.  The maps are in units of antenna temperature as measured at K-band for synchrotron and free-free emission, and at W-band for dust emission.  Figure \ref{fig:fgbeta} shows spectral index maps binned to lower resolution, where color indicates the spectral index and the size of the circle indicates the significance of the fit result at that location.  
Figure \ref{fig:chimap} shows the best $\chi^2_\nu$ value achieved at each pixel on the sky.

Almost regardless of foreground model the fit works quite well outside the Galactic plane, giving low $\chi^2_\nu$ and foreground maps that are in good agreement with the MEM templates \citep{hinshaw/etal:2007} and other works~\citep{eriksen/etal:prep}. Error maps for synchrotron and free-free emission have similar morphology due to the degeneracy between their amplitudes.

\begin{figure}
 \plotone{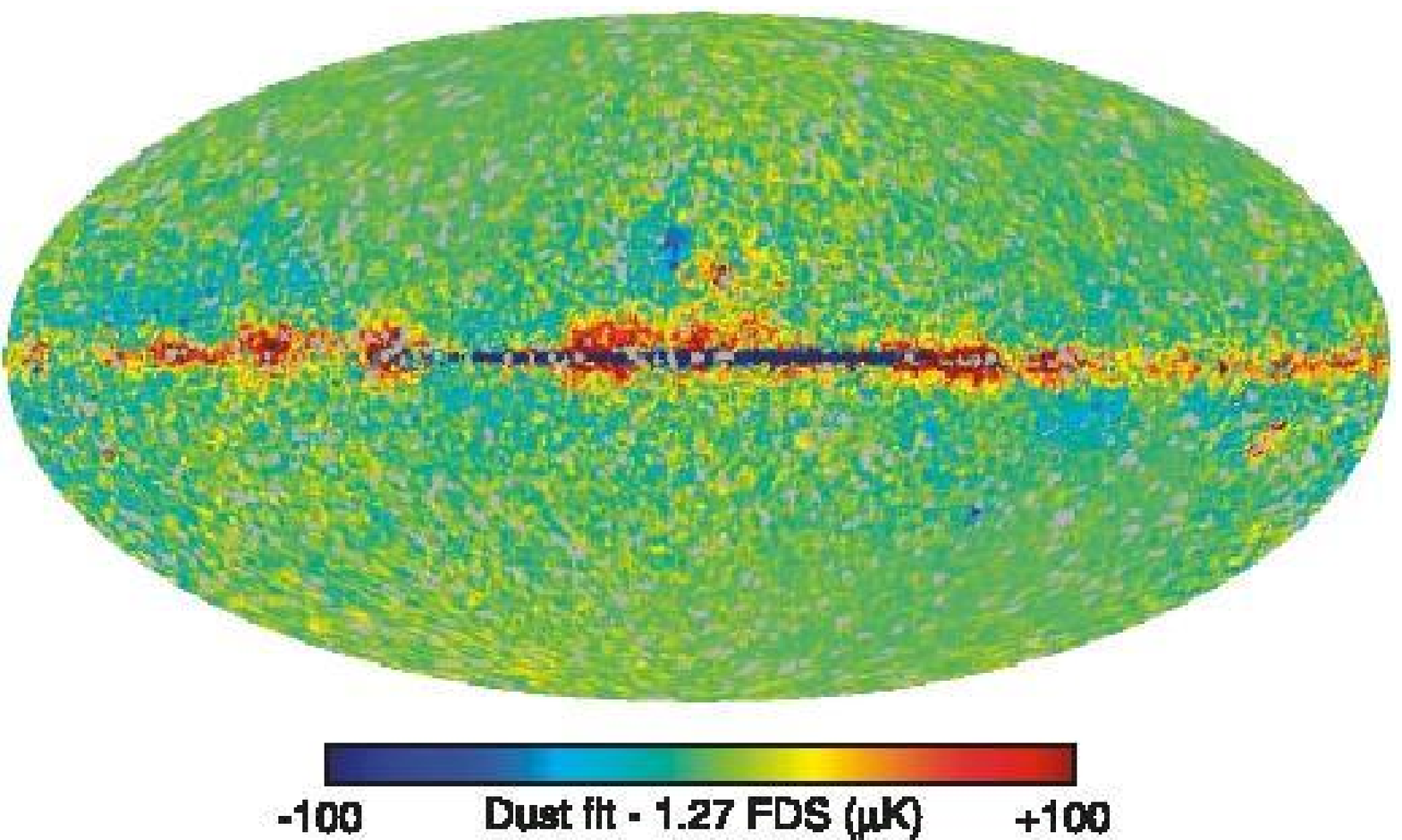}
 \caption{Residuals between the best-fit dust map from the ``base'' model and 1.27 times the ``model 8'' dust map of \cite{finkbeiner/davis/schlegel:1999} extrapolated to 94 GHz.  Pixels with $\chi_\nu^2 > 5$ or covered by the point source mask are colored gray.  
 The MCMC fit prefers less dust emission directly toward the Galactic center (deep blue pixels), and somewhat more emission further away (red and orange pixels).
 Units are antenna temperature at 94 GHz.\label{fig:dustcompare}}
\end{figure}

The overall dust brightness seems to be largely consistent with the template prediction \citep{finkbeiner/davis/schlegel:1999}, though the fit appears to prefer a spatial distribution somewhat less sharply peaked toward the Galactic center (Figure \ref{fig:dustcompare}).  The excess of observed emission compared to that predicted at 90 GHz, seen in the original model comparison with \textsl{COBE DMR} data, is still present.
Since the fit in the plane has high $\chi^2$ and is untrustworthy, the overall preferred spectral index for dust may be $< 2.0$, but the significance of this is not high.  Weighting with the covariance information from the fit and masking low-signal regions, the average value for $\beta_d$ in the ``base'' fit is $1.8$ with $\pm0.3$ from statistical errors and $\pm 0.2$ from systematic error depending on how the cuts are defined.   For comparison, model 8 of \cite{finkbeiner/davis/schlegel:1999} predicts $\beta_d = 1.55\pm0.01$ for a comparable sky cut.

The free-free component is consistent with expectations from previous fits and with H$\alpha$ observations when dust obscuration is taken into account.  
Free-free emission is quite high in the Galactic plane and in several regions (including Gum and Orion) appears to be dominant over synchrotron, even in K-band.
The ratio of the ``base'' fit free-free map to extinction-corrected H$\alpha$ map of \cite{finkbeiner:2003} was used to make a map of $h_{ff}$ (the temperature--H$\alpha$ intensity conversion factor).  A histogram was then made of all pixels with intensities larger than 5 Rayleighs (to mask out low-signal regions) and less than one magnitude of extinction (using the reddening map of \citealt{schlegel/finkbeiner/davis:1998}).  A gaussian fit to the peak of the histogram 
gives $h_{ff} = 11.8 \pm 8.8$ $\mu$K R$^{-1}$ at K-band, comparable to the value of 11.4 $\mu$K R$^{-1}$ expected from an electron temperature of 8000 K \citep{bennett/etal:2003c}, 
but also consistent with the lower values from template cleaning.

\begin{figure}
\plotone{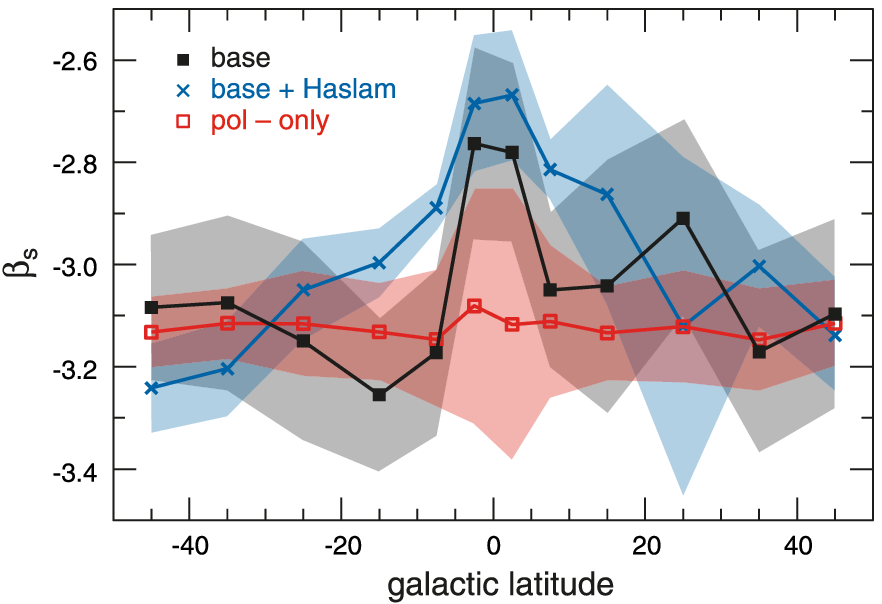}
 \caption{Synchrotron index plotted as a function of latitude for several fits. Pixels are binned by latitude, and only longitudes between 350 and 10 degrees are included.  Error regions indicate the 68\% scatter within each bin.  Solid black squares (gray region) are for the ``base'' model, blue crosses (blue region) are for the ``base'' model with 408 MHz data, and red empty squares (red region) are for a fit using only \WMAP's polarization data.  Fits where the dust spectral index was fixed to 1.7 and 2.0 are almost identical to the ``base'' fit.
 The trend to flatter spectral index in the plane does not appear when only polarization data are used, but the signal-to-noise is not high enough for the discrepancy to be significant.
 \label{fig:syncindex}}
\end{figure}

While subject to degeneracy with the free-free emission, synchrotron radiation is a stronger signal in \WMAP\ data than dust emission.
Pixel-by-pixel constraints become poor far from the plane, however there are still constraints on the best-fit spectral index.  For example, by comparing fits with constant spectral index, the Northern Polar Spur and the Fan region prefer an index of $-3.0$ or steeper.  All fits including total intensity data show the same preference for shallower spectral index in the plane as concluded by \cite{bennett/etal:2003c}; Figure \ref{fig:syncindex} shows  $\beta_s$ as a function of latitude for a number of different fits.  

From the polarized data, the synchrotron polarization fraction indicates strong depolarization toward the Galactic plane consistent with \cite{Kogut/etal:2007}.  Since Faraday rotation should not be large at these frequencies, this effect is due to multiple magnetic field orientations along the line of sight.
Dust polarization fraction appears to follow a pattern similar to the synchrotron polarization fraction, though the signal-to-noise ratio is low.  This is physically reasonable, as the polarization fraction is largely affected by the coherence of the magnetic field along the line of sight.  This implies that the dust intensity times a constant fraction may \emph{not} be the best template to use for dust polarization in the Galactic plane.

\subsection{The Galactic Plane}
Regions at very low latitudes are not as well fit by the ``base'' model, and there is dependence both on foreground
model and fit parameters.  A map of poorly fit regions reveals that they are in the brightest parts of the
Galaxy, where at these frequencies the free-free emission dominates.  

\begin{figure}
 \epsscale{0.9}
 \plotone{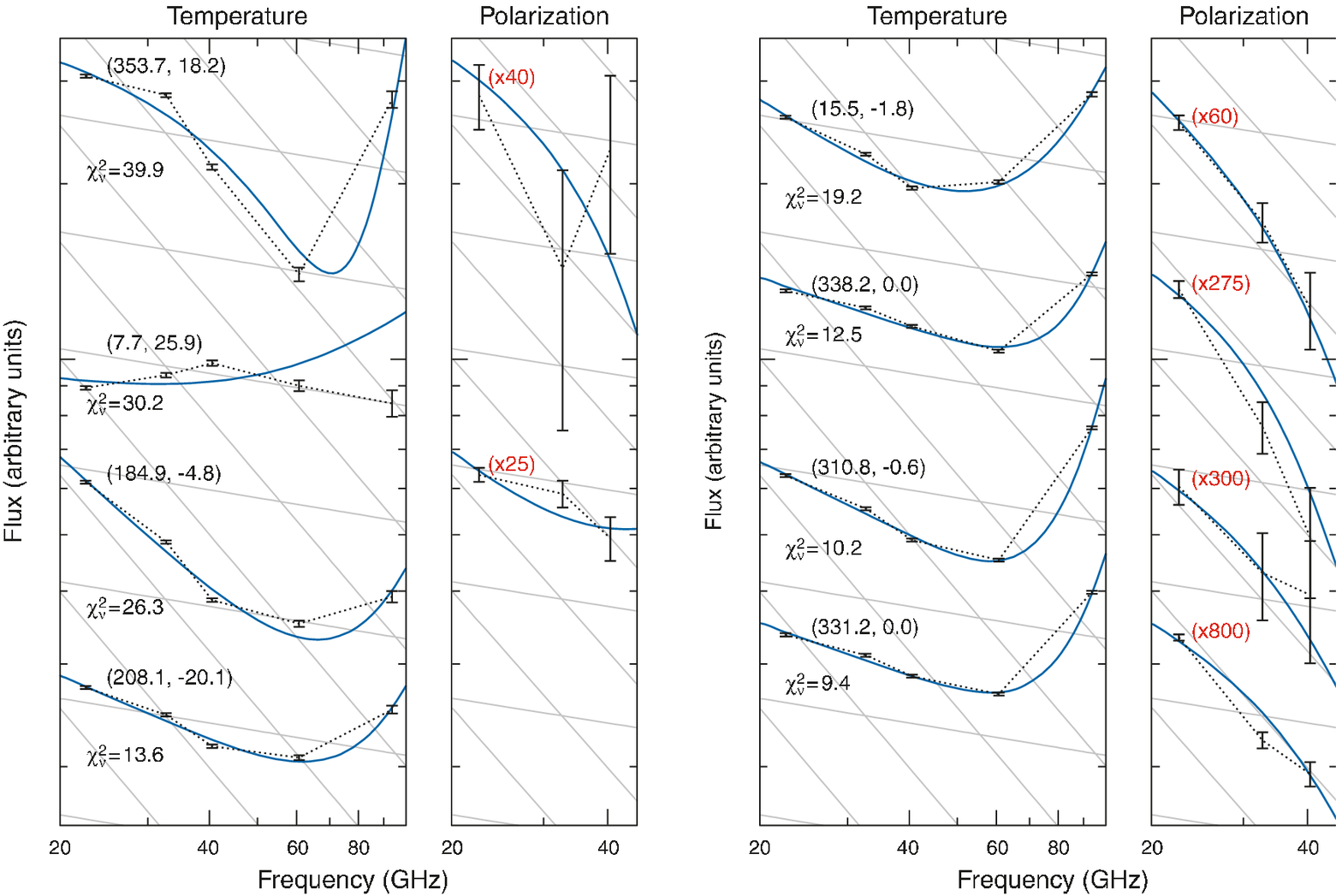}
 \caption{Temperature (Stokes $I$) and polarization ($\sqrt{Q^2 + U^2}$) spectra for poorly fit pixels in different regions of the sky.  The vertical scale is arbitrary {flux} units, but polarization data is shown beside the corresponding temperature data with a number in red indicating the approximate (K-band) ratio of intensity to polarization.  Polarization data is not shown for pixels with low signal to noise.  Numbers in parentheses are Galactic coordinates. The blue curve is the best-fit ``base'' model. The model does \emph{not} include synchrotron steepening; convex slopes are due to negative CMB contributions.  The grid of gray lines indicates spectral indexes of $\alpha = -0.14$ and $\alpha = -1$.  The left plot shows pixels further from the plane; the right plot shows pixels near the plane.  The fit converged for all regions shown and none were covered by the point source mask.  However, the four pixels shown on the left are (from top to bottom) within a few degrees of $\rho$ Oph, $\zeta$ Oph, Tau A, and the Orion nebula.
 \label{fig:funnypol}}
\end{figure}

\begin{figure}
 \epsscale{0.6}
 \plotone{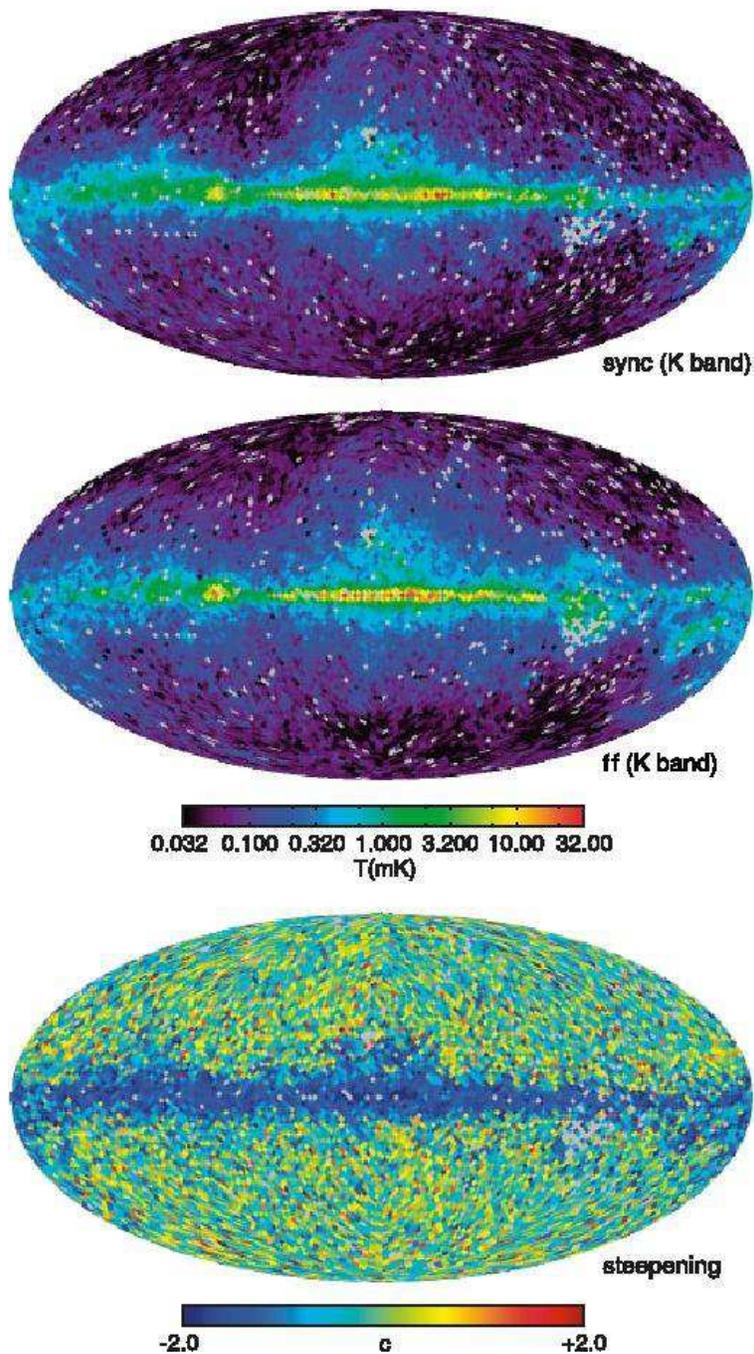}
 \caption{Maps for foreground components as determined by the MCMC fitting process for the ``steep'' model.  This model has a frequency dependent synchrotron spectral index $\beta_s (\nu) = \beta_s + c \ln (\nu/\nu_K)$.  Synchrotron and free-free temperatures are measured at K-band, and gray pixels are those masked due to point sources or flagged as problematic.  Since this model does not differ from the ``base'' model at high frequencies the thermal dust emission is unaffected and is not shown.  Note that the steepening parameter tends toward large negative values in the Galactic plane.  \emph{Top:} synchrotron; \emph{middle:} free-free; \emph{bottom:} steepening parameter $c$.
 \label{fig:fgsteep}}
\end{figure}

\begin{figure}
 \epsscale{0.85}
 \plotone{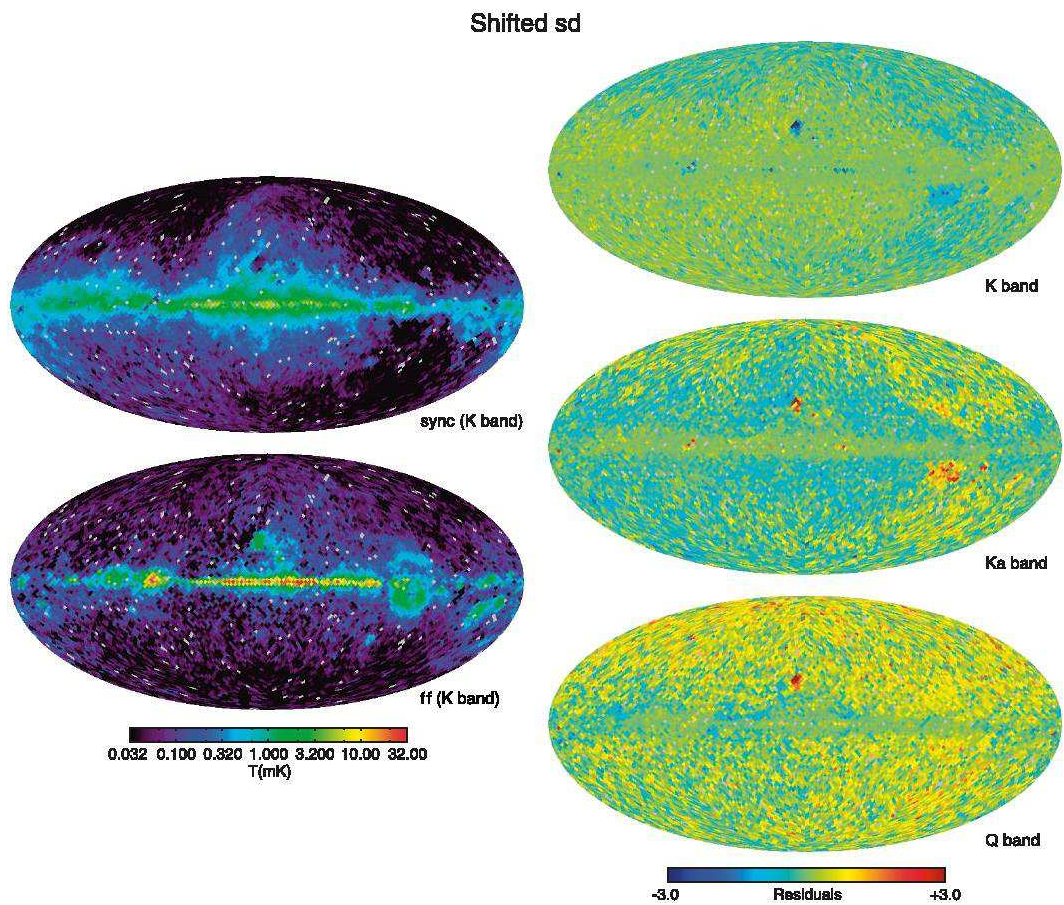}
 \caption{Foreground components and residual maps as determined by the MCMC fitting process for the ``shifted sd'' model.  This model adds a spinning dust component (not required by the fit) at low frequencies where the peak emission frequency has been lowered in an attempt to match the data.  Synchrotron and free-free temperatures are measured at K-band, and gray pixels are those masked due to point sources or flagged as problematic.  The spinning dust component of the fit is shown in Figure \ref{fig:tsdmap}.  This model does not differ from the ``base'' model at high frequencies and thus the thermal dust emission is unaffected and not shown here.  \emph{Left:}  synchrotron and free-free maps; \emph{right:} residuals (in dimensionless units of noise sigma) to the fit at K, Ka, and Q-bands.
 \label{fig:fgsd488}}
\end{figure}

\begin{figure}
 \epsscale{0.8}
 \plotone{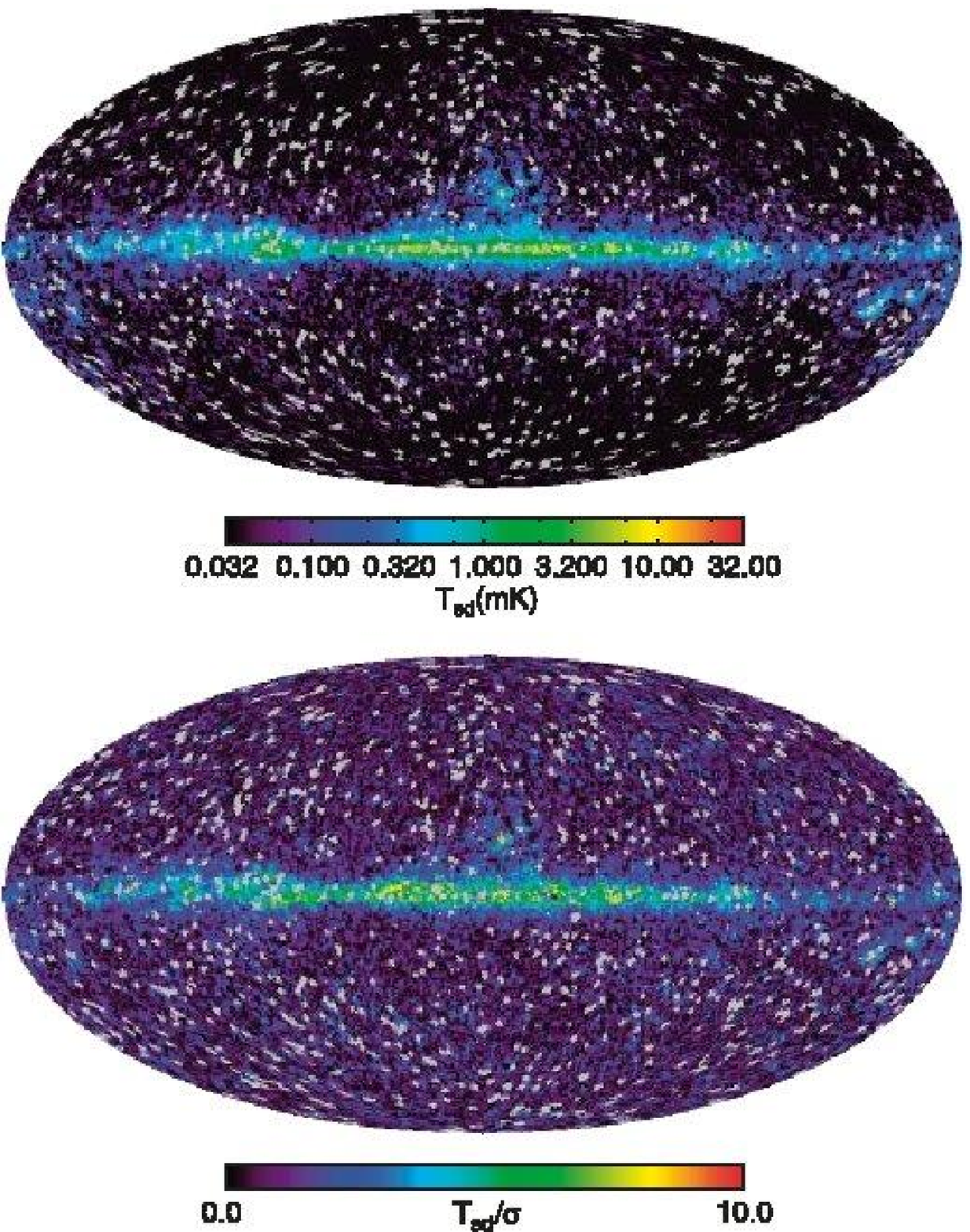}
 \caption{\emph{Top:} Map of possible spinning dust antenna temperature $T_{sd}$ (as measured at K-band) for the ``shifted'' spinning dust model fit.  While the model is based on a spinning dust spectrum, the data are insufficient for distinguishing the physical emission mechanism from other possibilities mentioned in the text. Maps of the other three foregrounds are qualitatively similar to that of the ``base'' model, with the synchrotron amplitude reduced somewhat to accommodate the additional low-frequency component.
 \emph{Bottom:} Map of spinning dust temperature divided by the marginalized temperature error as calculated by the MCMC chain.  This is statistical error only and does not include uncertainties in the model.
 \label{fig:tsdmap}}
\end{figure}

Pixels poorly fit by the ``base'' model have some common characteristics. Most are bright, but this is probably because similar less bright pixels have lower signal-to-noise and thus lower $\chi^2$. Many have a K-Ka temperature spectral index similar to what one would expect from free-free emission, but a considerably steeper Ka-Q spectral index.  Such pixels tend not to be highly polarized, and what polarized emission exists appears to be consistent with synchrotron emission with a typical spectral index of $\beta \approx -3$.
Data for several such individual pixels
are shown in Figure \ref{fig:funnypol}.

For the published cold neutral medium spinning dust model (the ``exact'' model of Table \ref{tab:chi}), the
maximum fraction of Ka-band flux attributable to spinning dust
is 17\% outside of the Galactic plane (using the 85\% mask).  The maximum full-sky
fraction of Ka-band flux attributable to spinning dust is 20\% for this model. However, this model
still does not provide a good fit within the Galactic plane ($\chi^2_\nu$ in this region is 1.63).

Allowing the spinning dust spectrum to shift in frequency to obtain a better fit results
in a Ka-band flux fraction of 14\% for spinning dust, roughly independent of sky cut.
A map of the spinning dust component from this fit and its error is shown in Figure \ref{fig:tsdmap}.
The morphology lies somewhat between that of dust and free-free emission, though the details depend on the specifics of the model.
The Galactic plane is equally well-fit by adding a synchrotron steepening parameter $c$ into the fit.  The actual value of $c$ is generally not well-constrained, but the average value in the plane is $-1.8$. This very rapid steepening does not appear to be consistent with cosmic ray models \citep{strong/moskalenko/ptuskin:2007}, but may have some other physical origin.

Figure \ref{fig:fgsteep} shows the low-frequency foregrounds given by the MCMC fit using the ``steep'' model.  Thermal dust emission is indistinguishable from the the ``base'' fit.  Residual maps from this fit are featureless, as hinted at from the $\chi^2$ information in Table \ref{tab:chi}.  Figure \ref{fig:fgsd488} shows low-frequency foregrounds and residuals in K, Ka, and Q-bands for the ``shifted sd'' fit which includes a spinning dust-like component.  This model produces a good fit in the plane, but seems to have some problems with the Ophiuchus and Gum regions.

Since the goodness-of-fit outside the plane
is not improved by the addition of a spinning dust component,
and low signal-to-noise regions bias the spinning dust fraction upwards, we
regard the spinning dust fraction of the fits above as an upper bound to the overall amount of diffuse
spinning dust emission present.  As with previous \WMAP\ fits,
this new fitting technique continues to find that spinning dust is a
subdominant emission process.

\section{Discussion\label{sec:discussion}}

\subsection{Effect on CMB and Cosmology}
The uncertainties of the fit in the Galactic plane preclude CMB analysis for those regions.  Fortunately, such regions appear to be tightly confined to the plane inside a very narrow sky cut (9\% of the total sky) and thus can be excluded without losing much information for cosmological analysis.  The foreground maps from the MCMC fit are similar to those from the MEM fit and other foreground templates, which means CMB polarization maps cleaned using such templates will also be similar.

\begin{figure}
\epsscale{1.0}
\plotone{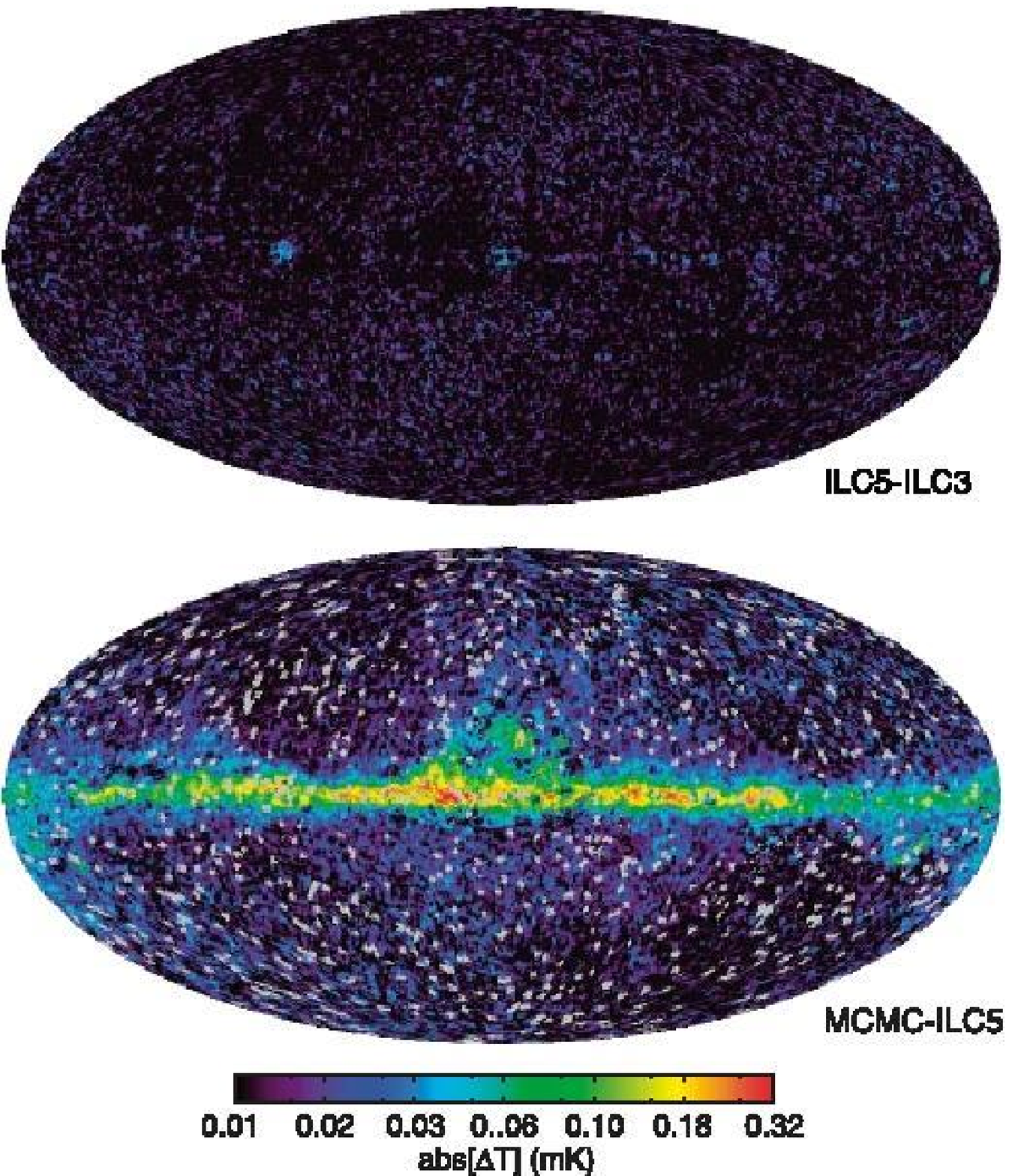}
 \caption{Difference maps between the five-year and three-year ILC maps (top), and the MCMC fit for the CMB and the 5-year ILC map (bottom). For the former, the most prominent large-scale difference is from the reduction of noise with \WMAP's observational pattern.  Even for the latter, the variance outside the KQ85 analysis mask is 116 $\mu$K$^2$, lower than the CMB power at large scales.  \label{fig:ilcdiff}}
\end{figure}

Outside of a narrow band on the Galactic plane the CMB map produced by the fit is visually identical to the ILC map.  The difference between the ``base'' fit CMB map and the ILC map is shown in Figure \ref{fig:ilcdiff}.  The total variance of this difference map outside of the KQ85 mask used for power spectrum analysis is 116 $\mu$K$^2$, much lower than the CMB power.  The variance between the ``base'' fit CMB map and the ``shifted sd'' CMB map is 44.1 $\mu$K$^2$ outside of the KQ85 mask; variance from one fit to another is generally even smaller for other combinations.
Spherical harmonic decomposition did not show these total variances to be strongly focused at any particular multipole,  and the numbers are small enough that differences between maps fall within cosmic variance.

The CMB polarization maps produced by the MCMC fit presented in this work are noisy and show some evidence of synchrotron contamination.
Nevertheless, the covariance maps from the fits can be used to bound the amount of contamination present, and are available on the LAMBDA website.  These are produced from the (marginalized) variance of each parameter over the Markov chain for each pixel.  For cosmological analysis a different method is used to marginalize over polarization foregrounds. For a full description see \cite{dunkley/etal:prep}.

\subsection{More Complicated Models}
All of the models so far fit assume that the spectral shape of foreground emission in a $\sim\! 1 \textrm{deg}^2$ pixel can be described as a sum of power-laws or other simple shapes.  This is justified if the observed emission is dominated by a few emission mechanisms which simply combine additively along the line of sight and have minor spatial variation within the beam.  In more complex regions of the Galaxy, however, things may not be so simple.

If two synchrotron regions along the line of sight have their polarization angles oriented nearly orthogonally, then the total polarized emission will be sharply reduced. If the two regions have different spectral indices then the cancellation in polarization will be maximized at the frequency where the individual polarization amplitudes match, causing a dip in the polarization spectrum.  Thus even with pure synchrotron emission the polarization spectrum can look quite different from the temperature spectrum. 

To assess this effect 100,000 Monte Carlo realizations were made of a superposition of two independent randomly oriented synchrotron emitting regions.  Parameters for the distribution of intensity and spectral index were chosen to roughly correspond to observations, but the simulation was meant only to provide a generous estimate of how different temperature and polarization behavior could be simply due to multiple synchrotron regions along the line of sight.  The mean spectral index difference was small ($-0.051$) but the standard deviation was not insignificant ($0.12$) and the distribution was non-gaussian with high kurtosis ($4.4$).  Over $20\%$ of the simulations had an absolute slope difference larger than $0.1$.  Thus we cautiously conclude that while differences between temperature and polarization spectral indices at the $\sim\!\! 0.1$ level could be quite mundane in origin, consistent differences of $0.25$ or larger are probably \emph{not} due to chance alignments in polarization angle and may be caused by an unpolarized non-thermal temperature component. 

For the fit, free-free emission was modeled as a pure power law based on the assumption that the plasma is optically thin.  In reality, {\hii} regions can become dense enough to become optically thick at frequencies as high as 20 GHz, although such regions are spatially small and do not contribute significantly to the observed emission for a beam as large as \WMAP's.
Further, to obtain rising flux at K-band requires very high emission measure ($\sim\!\! 10^9$ cm$^{-6}$pc).
Even if they were somehow significant, such regions can not explain a steepening spectrum past Ka-band.

Synchrotron self-absorption can also cause a low-frequency turnover, but the physical parameters necessary for the turnover frequency to lie in or near the \WMAP\ range imply conditions typically only
found in active galactic nuclei or other extreme regions.  It may be physically possible for synchrotron radiation from stellar-mass black hole jets or accretion disks to become optically thick at \WMAP\ frequencies, but this phenomenon has yet to be clearly observed and it is unlikely that such emission would contribute significantly at \WMAP's resolution. 

Synchrotron radiation is suppressed when emitted from a region with a refractive index less than unity, such as a plasma.  This is known as the Tsytovitch-Razin effect, and causes strong suppression of synchrotron emission below 
$20(n_e/B)$ Hz, where $n_e$ is the electron density (in cm$^{-3}$) and $B$ the perpendicular component of the magnetic field (in G).
For typical Galactic electron densities and magnetic fields, this cutoff is in the 3--300 MHz range, at most.
Unless energetic electrons play an unexpectedly large role in diffuse emission, \WMAP\ should not see significant Tsytovitch-Razin suppression.

Diffusive synchrotron radiation (DSR) differs from an ideal synchrotron spectrum because of the presence of significant random fluctuations in the magnetic field~\citep{2005astro.ph.10317F}.  In this model 
lower-energy electrons experience small-scale turbulence in the magnetic field structure and follow non-circular paths due to the random deflections.
In such models the emission spectrum can turn over from a power law with $\beta \approx -2.1$ in the turbulence-dominated diffusive regime to a normal synchrotron spectrum at higher frequencies.  While in most models this occurs at low frequencies, there is some indication that for pulsar wind nebula the turbulence is relevant up to the GHz range and above \citep{fleishman/bietenholz:2007}.
Whether DSR can occur for less compact objects is not understood at this time. 

\subsection{Other Components}

Much has been written on the possible presence of anomalous emission in the lower frequency bands of \WMAP. There are at least two categories of observations: one is of emission that is diffuse and significant over large portions of the sky outside the Galactic plane, and another where the emission is important and perhaps even dominant in specific compact regions.

Much evidence for diffuse anomalous emission comes from template correlations \citep{deoliveira-costa/etal:1999, bonaldi/etal:2007, dobler/finkbeiner:2007} rather than direct fitting of the data (though for a recent example of the latter, see \citealt{miville-deschenes/etal:prep}).  
 Characterizing the error and offsets in templates made from data at very different frequencies has proven challenging. 
Nevertheless, \cite{dobler/finkbeiner:2007} show that using an H$\alpha$ template to fit \WMAP\ data results in an improvement of the $\chi^2_\nu$ by 0.016 (to 2.977), and that the spectrum has a significant ``bump''.  Though the improvement seems small, due to the large number of degrees of freedom it is statistically significant, and appears to be robust against the systematic error investigated in that work. The Ka-band excess for select Galactic plane pixels (Figure \ref{fig:funnypol}), however, is in the data alone, independent of any template.  \cite{boughn/pober:2007} also find from combining 19 GHz data with \WMAP\ K-band that the Galactic plane seems to have antenna temperature falling less steeply than $\beta = -2$ (i.e. a rising flux spectrum).

Specifically regarding \emph{compact} regions, \citet{finkbeiner/etal:2002} previously reported on two regions which might show excess emission in the 10--40 GHz range due to spinning dust.  CBI observations \citep{2006ApJ...643L.111D} failed to find anomalous emission from one, LPH96 201.663+1.643; other authors \citep{mccullough/chen:2002} had previously raised the possibility that such emission might be due to an optically thick ultracompact {\hii} region.  The other, LDN 1622, was found by \cite{2006ApJ...639..951C} to have a spectral energy distribution consistent with spinning dust, a result driven in part by the lack of flux at 5 GHz found by the Parkes-MIT-NRAO survey of~\cite{1993AJ....106.1095C}.  Other surveys at lower frequencies with a larger angular resolution more comparable to \WMAP's, however, have not measured a lack of flux -- the 408 MHz data summarized by \cite{haslam/etal:1981} measures more flux than \WMAP\ K-band and is consistent with a mixture of power-law thermal and non-thermal components for the region containing LDN 1622.

Separately, \cite{Scaife:2007mz} recently observed a sample of northern {\hii} regions and found no evidence for anomalous emission in any, but observations with the Very Small Array \citep{Scaife/etal:2007b} find some evidence for a 33 GHz excess in SNR 3C396. Thus for compact regions the status of anomalous emission appears to be mixed.

There has been some discussion in the literature of correlation between CMB maps and neutral hydrogen \citep{2007arXiv0704.1125V}, but this result was not found to be statistically significant \citep{2007arXiv0706.1703L}. 

\subsection{Directions for the Future}
It is quite probable that at least one of the above model complications or additional components is relevant for understanding our Galaxy.  More data is needed, particularly in the 5 to 30 GHz range.  Further, the inability to measure flux at large angular scales is a problem for many observations, particularly when the angular scale limit depends on the observing frequency.  This continues to make precise comparison of results difficult.  Large-scale observations with calibration errors at the percent level or better are needed to address the nature of features seen in some pixels of the Galactic plane.  

For the \WMAP\ foreground fits the dust spectrum was treated as a pure power law.  In reality, dust emission in the \WMAP\ bands is probably dominated by a cold component with a low enough temperature that the exponential cutoff is not negligible.  However, since the frequency range of \WMAP's dust sensitivity is narrow the largest effect of the exponential correction is simply a modification of the apparent power-law index, which for typical cold dust temperatures ($\sim\!\!10$ K) amounts to a change of about $0.1$ in $\beta_d$.  This effective bias is therefore small compared to typical errorbars.  
However, there are already hints that the extrapolation of dust models to millimeter wavelengths is not entirely satisfactory.  

Further insight on Galactic foregrounds will be obtained from upcoming experiments.
For example, the Planck satellite \citep{tauber:2005}, scheduled to launched in 2008, will soon provide more insight on Galactic foregrounds.  While Planck's frequency coverage does not extend low enough to overlap \WMAP's K-band, Planck will observe at \WMAP's other frequencies with roughly 25\% narrower beams and an order of magnitude better sensitivity.  Comparison to \WMAP's results will be an important check of systematic errors, and the increased sensitivity can help with foreground discrimination, for example by improving knowledge of the spectral index for polarized synchrotron emission.
Further, Planck has six higher frequency channels in the 100--860 GHz range, which will be invaluable for studying dust to a precision several orders of magnitude better than what was available with prior data in this frequency range.


\section{Conclusions}
\begin{itemize}
\item \WMAP's temperature and polarization data outside the Galactic plane are well described by the standard three foreground components: synchrotron, free-free, and thermal dust, each with power-law spectral indices.
\item The spectral index for synchrotron radiation at high latitudes is consistent with $\beta \approx -3$, with trend toward $\beta = -2.7$ seen at lower latitudes.  The spectral index for dust is not well constrained but appears consistent with $\beta \cong 2$.
\item Some localized regions in the Galactic plane show emission with $\beta\approx -2$ below $33$ GHz which steepens by as much as $\Delta\beta = -0.8$ above $33$ GHz, and this emission is mostly unpolarized.  Both spinning dust and synchrotron steepening models can be used to fit this emission component, whose physical origin is unclear.
\item CMB maps from different model fits show $< 50$ $\mu$K$^2$ of variance relative to each other outside the KQ85 analysis mask, and $< 120$ $\mu$K$^2$ of variance relative to the five-year ILC map. The CMB and cosmological results are robust to changes in the foreground model.
\item \WMAP\ serves as a precise ($< 1\%$ error), unbiased, full-sky survey of the Galaxy which can  reveal large-scale microwave emission features never before seen.
\end{itemize}

\acknowledgements
The \WMAP\ mission is made possible by the support of the Science Mission Directorate Office at NASA Headquarters.  This research was additionally supported by NASA grants NNG05GE76G, NNX07AL75G S01, LTSA03-000-0090, ATPNNG04GK55G, and ADP03-0000-092.  This research has made use of NASA's Astrophysics Data System Bibliographic Services.  We acknowledge use of the HEALPix, CAMB, and CMBFAST packages.



\end{document}